\documentstyle[emulateapj]{article}
\newcommand{\preprint}[1]{#1}

\newcommand{\url}[1]{{\tt #1}}

\slugcomment{To be published in the Astronomical Journal, 2001 April 1}

\lefthead{Helfand {\it et al.}}
\righthead{Long-Term Quasar Variability}

\begin{document}

\title{Long-term Optical Variability of Radio-Selected
Quasars from the {\it FIRST} Survey}

\author{David J. Helfand}
\affil{Astronomy Dept., Columbia University, New York, NY~10027,
djh@astro.columbia.edu}
\authoremail{djh@astro.columbia.edu}

\author{Remington P.S. Stone}
\affil{Lick Observatory, Mt Hamilton, CA~95140
rem@ucolick.org}
\authoremail{rem@ucolick.org}

\author{Beth Willman}
\affil{Department of Astronomy, University of Washington, Seattle,
WA~98195-1580, willman@orca.astro.washington.edu}
\authoremail{willman@orca.astro.washington.edu}

\author{Richard L. White}
\affil{Space Telescope Science Institute,
3700 San Martin Dr., Baltimore, MD 21218, rlw@stsci.edu}
\authoremail{rlw@stsci.edu}

\author{Robert H. Becker}
\affil{Physics Dept., University of California, Davis, CA  95616\\
and IGPP/Lawrence Livermore National Laboratory, bob@igpp.llnl.gov}
\authoremail{bob@igpp.llnl.gov}

\author{Trevor Price \&  Michael D. Gregg}
\affil{Physics Dept., University of California, Davis, CA  95616}
\authoremail{gregg@igpp.llnl.gov}

\author{Richard G. McMahon}
\affil{Institute of Astronomy, Madingley Road, Cambridge CB3 0HA, UK}
\authoremail{rgm@ast.cam.ac.uk}

\begin{abstract}

We have obtained single-epoch optical photometry for 201 quasars, taken
from the {\it FIRST} Bright Quasar Survey, which span a wide range in
radio loudness. Comparison with the magnitudes of these objects on the
POSS-I plates provides by far the largest sample of long-term
variability amplitudes for radio-selected quasars yet produced. We find
the quasars to be more variable in the blue than in the red band,
consistent with work on optically selected samples. The
previously noted trend of decreasing variability with increasing
optical luminosity applies only to radio-quiet objects.  Furthermore,
we do not confirm a rise in variability amplitude with redshift, nor do
we see any dependence on radio flux or luminosity. The variability over
a radio-optical flux ratio range spanning a factor of 60,000 from
radio-quiet to extreme radio-loud objects is largely constant, although
there is a suggestion of greater variability in the extreme radio-loud
objects.  We demonstrate the importance of Malmquist bias in
variability studies, and develop a procedure to correct for the bias in
order to reveal the underlying variability properties of the sample.

\end{abstract}


\section{Introduction}

Variability on timescales ranging from hours to decades has been one of the
defining characteristics of active galactic nuclei (AGN) since their discovery
nearly forty years ago. These changes in the source flux received at
Earth can arise from many causes: variability in the source's central
engine, changes in a relativistic beam's orientation or velocity, changes
in absorption along the line of sight to the active nucleus, gravitational
microlensing, and interplanetary or interstellar scintillation. As
such, AGN variability provides an important diagnostic in studies of the
physics of the central engine, the nuclear environment, the properties
of the material along the line of sight, and, ultimately, of the evolution of
the AGN population.

Twenty-five years ago, Grandi and Tifft (1974) published the first
compilation of quasar variability studies. Citing 92 references to
previous work, they listed photographic and photoelectric photometry
for a heterogeneous sample of 86 bright ($B<18$) quasars, concluding
that typical uncertainties in quasar optical magnitudes resulting from
long-term variability were $\sim0.25$ magnitudes.  A conference devoted
exclusively to AGN variability -- on all timescales and throughout the
electromagnetic spectrum -- summarized the status of the subject at the
beginning of the last decade (Miller and Wiita 1991). Since that time,
over a dozen major studies of optical variability have appeared
(Table~1), covering timespans up to thirty years and including over
1000 objects. Most, however, include only optically selected quasars;
Smith et al. (1993), Netzer et al.  (1996), Sirola et al. (1998), and
Garcia et al. (1999) are the exceptions with a combined total of 160
radio-loud quasars in their samples.  Considerable disagreement remains
regarding the correlation of variability amplitude with quasar power,
redshift, and other quasar properties (see the references in Table~1).

%
%
\begingroup
\begin{deluxetable}{cccccccc}
\tablewidth{0pt}
\tablecaption{Summary of Long-term Quasar Optical Variability Studies}
\tablenum{1}
\tablehead{
\colhead{Paper} & \colhead{Maximum} & \colhead{Sample} & \colhead{Colors} & \colhead{Number of} & \colhead{Number of} & \colhead{Number of Radio-}\\
\colhead{(date)} & \colhead{Timescale} & & & \colhead{Objects} & \colhead{Observations} & \colhead{Selected Objects}\\
 & \colhead{(yrs)}
}
\startdata
Netzer \& Sheffar &            31  &     84 deg$^2$  &      O  &          64  &         2  &             0\\
(1983)\\
Cristiani et al.  & 7  &       SA 94 &        B  &          90 &         15  &             0\\
(1990)\\
Giallongo et al.  & 4-11 &  Braccesi + SA57  &  B &          55 &         7-9  &             0\\
(1991)\\
Ciamati et al.  &    2 &      small area  &  U,F,J  &        52  &         3  &             0\\
(1993)\\
Smith et al. &       21 &      radio-loud   &  B, Ph &       60 &         40-100 &          60\\
(1993)\\
Hook et al. &       16 &         SGP &        B$_{\rm J}$ &        300  &        12  &             0\\
(1994)\\
Trevese et al.  &   15 &        SA 57  &       B$_{\rm J}$ &          35  &        11 &              0\\
(1994)\\
Hawkins & 15 &        south &      IIIA$_{\rm J}$ &         71  &      $\sim 10$ &             few\\
(1996)\\
Cristiani et al. &          5 &        SA 94 &         B &         180  &        10   &            0\\
(1996)\\
Netzer et al. &      6  &       north &         B  &         44  &       25-50 &           44\\
(1996)\\
Cristiani et al. &          8 &        SA 94 &        B,R &        149  &         8  &             0\\
(1997)\\
Trevese et al.  &    6 &        SA 57 &      U,B,F  &       26  &        24 &              0\\
(1997)\\ 
Sirola et al. &            4-7 &     Las Campanas &   V,R  &       151 &       5-10 &                         21\\
(1998)\\
Garcia et al.  &     3  &       V$<16.5$ &    V  &                 50  &      5-10 &             35\\
(1999)\\
Giveon et al. &      7 &         PG &          B,R &         42  &      30-60 &            0\\
(1999)\\
This work &         48 &        FBQS  &        B,R  &       202 &          2 &            202\\
\enddata
\end{deluxetable}
\endgroup


We are in the process of compiling a very large sample of bright
quasars using the Very Large Array (VLA)\footnote {The VLA is part of
the National Radio Astronomy Observatory which is operated by
Associated Universities, Inc., under cooperative agreement with the
National Science Foundation.} {\it FIRST} survey to select candidates
from the POSS-I plates for optical spectroscopic followup. As part of
this program, we have obtained current photometric $B$ and $R$
magnitudes for over 200 confirmed quasars, providing variability
estimates on proper timescales of 10--45 years for a large quasar
sample spanning a broad range in radio loudness. In section 2, we
briefly describe the {\it FIRST} quasar survey and the selection from
it of our variability sample. We go on to describe the photometric
observations and their reduction (\S3) as well as the procedures used
to calibrate the photometry of the archival POSS-I data (\S4). Section
5 presents our results including a comparison with previous work and an
analysis of the dependence of variability amplitudes on radio loudness,
redshift and luminosity, as well as the important effect Malmquist bias
has on the distribution of the magnitude variations. We conclude with a
summary of our results.

\section{The {\it FIRST} Bright Quasar Survey}

The {\it FIRST} Bright Quasar Survey (FBQS) has recently been described
in detail by White et al. (2000). Briefly, the survey is based on a
comparison of the radio catalog from the VLA {\it FIRST} survey (Becker
et al. 1995; White et al. 1997) and the Cambridge Automated Plate
Machine scans of the Palomar Observatory Sky Survey-I (POSS-I) plates
(APM -- McMahon and Irwin 1992; McMahon et al. 2000). All optical
objects brighter than $E$ (red) magnitude 17.8 (after corrections for
extinction)\footnote{The initial selection for the FBQS came from the
APM magnitudes. Subsequent calibration of these magnitudes using the
APS scans (see McMahon et al. 2000 for details) led to a small number
of objects already spectroscopically confirmed as quasars being dropped
from the sample presented in White et al. (2000) because their revised
POSS-I magnitudes fell slightly below the $R=17.8$ threshold. This
program, however, was begun before the recalibration. Thus twenty-one
of the sources observed have POSS-I magnitudes slightly fainter than
$R=17.8$ as can be seen in column 9 in Table 2. Eight of these have
been published in Gregg et al. (1996), while the remainder appear here
for the first time.} classified as stellar on at least one of the two
POSS-I plates, which are bluer than $O-E=2.0$ and lie within
$1.2^{\prime\prime}$ of a radio source, are selected as quasar
candidates. Spectroscopic followup has demonstrated that over 55\% of
these objects are quasars, with another $\sim$10\% classified as BL Lac
objects; the remaining sources are a mix of narrow-line AGN, HII
galaxies, and radio galaxies lacking emission lines. Over 700 quasars
and BL Lacs have been found in the first $\sim$2700 deg$^2$ of the
survey; $\sim75$\% are newly discovered objects.

For our followup program of optical photometry, 201 quasars were
selected at random from both the portion of the north Galactic cap
survey reported in White et al. (2000) and its extension to the {\it
FIRST} equatorial strip in the south Galactic cap (Becker et al.
2001).  Suitability for the Right Ascension range of each scheduled
observing run was the only selection criterion; the magnitude, radio
flux density, and redshift distributions of the chosen quasars are
statistically identical to these distributions for the FBQS sample as a
whole. Flux densities for the sample cover a range 1.0 - 15,000 mJy,
and objects with redshifts from 0.1 to 3.4 are included. The ratio of 5
GHz to 2500~\AA\ flux densities $R^*$ used to characterize the radio
loudness of an object (Stocke et al. 1991) spans the range 0.3 to
22,000; objects with $R^{*}<10$, generally characterized as radio-quiet
quasars, constitute 46\% of the sample.

\section{Optical Photometry}

Observations of the selected objects were carried out by one of us
(RPSS) in a number of observing runs at the Lick 1-m telescope over the
period 20 December 1995 through 6 June 1997. The 2048 $\times$ 2048
pixel Orbit CCD provides a $6^{\prime}$ field of view, sufficient to
include a number of stars in each image for use in calibrating the APM
magnitudes. Observations were conducted through standard $B$ and $R$
filters; typical integration times were 100--600~s divided into two
frames to facilitate cosmic ray rejection.  Standard stars from the
list of Landolt (1992) were observed at least twice per night to solve
for the extinction correction. Conditions during the observations
ranged from photometric to light cirrus.

The data were reduced using standard IRAF routines for flat-fielding,
cosmic ray removal, calibration and photometry. In particular, the IRAF
task PHOT was used to perform aperture photometry on both the quasar
target and all other stellar objects of similar magnitude in the
field.

\section{Calibration of the APM Magnitudes}

The preliminary photometric calibration of the APM POSS-I scans
currently in use is described in some detail in McMahon et al. (2000).
For stellar objects, the global rms photometric uncertainty is 0.5
magnitudes in the range of interest here ($15<O,E<19$), a precision
insufficient for detecting the expected variability amplitudes for most
quasars. Thus, we began by using our CCD photometry in the 201 quasar
fields to calibrate each of the 53 POSS-I plates on which our objects
were found.

There are from 1 to 15 quasar fields per plate. In each field, we
selected isolated stellar objects with $13<B,R<19$ for use in forming a
sample for photometric calibration. The magnitudes in the two CCD
exposures for each star were compared; if they differed by more than
0.2 magnitudes, the star was discarded. The stars were then matched to
the APM catalog with a matching criterion of $10^{\prime\prime}$ (in
order to account for proper motion over the $\sim45$-year interval
between the POSS-I and our observations); any star with no match was
discarded. If two stars matched within this radius, the nearer one was
chosen unless it differed in brightness by more than 2 magnitudes.
These procedures yielded a database of 1695 stars.

We then performed an interative, linear least-squares fit to the
magnitude differences $(B-O, R-E)$ as a function of $O,E$ magnitude in
order to derive a magnitude-dependent calibration for each pair of
plates; stars lying more than $3\sigma$ from the best fit on either
plate were deleted from the database in order to eliminate variable
stars, mismatched stars, images affected by cosmic ray hits, etc. We
required a minimum of five stars per plate for an acceptable
calibration; the median was 16 stars per plate, with the best covered
regions having more than 80 stars per plate. The mean (and median) rms
values from the fits to both sets of plates are 0.15 magnitudes; the
fit for the POSS-I $O$ plate 1146 is displayed in
Figure~\ref{figcalib}. The corrections were then applied to the $O$ and
$E$ magnitudes for each quasar.  We estimate that the combined uncertainty
in the magnitude differences due to measurement and calibration errors
is 0.17 magnitudes for both the red and blue.

\placefigure{figcalib}

\preprint{
\begin{figure*}
\centering \leavevmode
\epsfxsize=0.6\textwidth \epsfbox{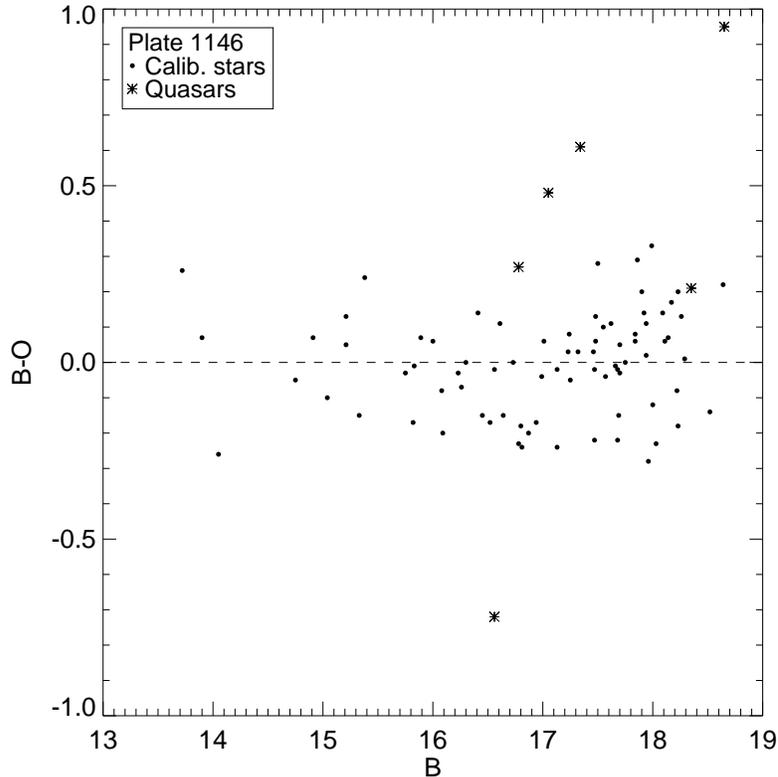}
\caption{
The calibration results for the POSS-I $O$ plate 1146. Stars selected
from our CCD images for use in the calibration are represented by dots,
while quasars are shown as asterisks. The rms for the 78 stars is 0.14 mag; note
the substantially higher scatter for the quasars, as well as the marked
tendency for the quasars to be fainter (positive $B-O$) at the current epoch.
}
\label{figcalib}
\end{figure*}
}

As an additional check on the calibration procedure, we compared our
CCD-corrected magnitudes with the APS-corrected magnitudes which we
have derived for the entire APM/{\it FIRST} identification catalog
described in McMahon et al. (2000). The distribution of corrections was
very similar; since the scatter in the APS magnitudes was slightly
larger, we used our CCD plate-by-plate calibrations when they were
available. For eight of our quasars, there were too few calibration
stars per plate, and the APS magnitudes were used; these sources are
flagged in Table 2.

\section{Long-term Quasar Variability}

\subsection {Photometric results}

In Table 2, we present the calibrated $O$, $B$, $R$, and $E$ magnitudes
for our 201 quasars. Each source's {\it FIRST} coordinates are given in
the first two columns. The next two columns list the dates of
observation for the POSS I plates (both colors were taken on the same
night) and for our CCD observations, followed by the calibrated
magnitudes and $B-O, R-E$ variations (positive $B-O, R-E$ values mean
the source is fainter now than when it was first detected on the POSS I
plates). We also include the {\it FIRST} peak and integrated 20cm flux
densities, the measured redshift (from White et al. (2000), Becker et
al. (2001), Gregg et al. (1996) and our own unpublished values for the
remaining objects with $E>17.8$), and values for the absolute blue
magnitude, radio luminosity, and $\log R^*$, the radio loudness
parameter; an $H_o = 50$ km s$^{-1}$ Mpc$^{-1}$, $\Omega = 1$, $\Lambda
= 0$ cosmology is assumed throughout. Note that the radio luminosities
are for the radio core {\it only} and do not include flux from any
extended lobes. The final column contains a radio morphology flag;
about 10\% of the sources are classical double-lobed sources, while
most of the remainder are point-like at the $\sim5^{\prime\prime}$
resolution of the {\it FIRST} survey.

By far the most variable object in the sample is the EGRET source
4C38.41 (FIRST J131059.4+323334), a well-known Optically Violent Variable which was 4.45 and
4.17 magnitudes brighter in the $E$ and $O$ bands, respectively, in
1950 than when we observed it in 1996. Since Barbieri et al. (1977) saw
this object change from B=15.85 to B=19 over the course of four years,
the large variation we report is not unexpected. The only other object
which varied by more than 1.5 magnitudes in either band is the
low-redshift ($z=0.28$), core-dominated, flat-spectrum quasar B1719+357
for which $B-O=1.98$.

In all, 30 objects (15\%) varied by at least a factor of 2 in at least
one band.  Of these, only four have been reported previously in the
literature as optical variables. For the sample as a whole, there is a
large bias toward objects having faded since their POSS-I detections; a
detailed analysis of this effect, a manifestation of Malmquist bias, is
given below, following which we present the results of comparing
long-term variability with other quasar properties.

\subsection{Skewness of the $\delta m$ distribution}

\placefigure{figobsdist}

\preprint{
\begin{figure*}
\centering \leavevmode
\epsfxsize=0.6\textwidth \epsfbox{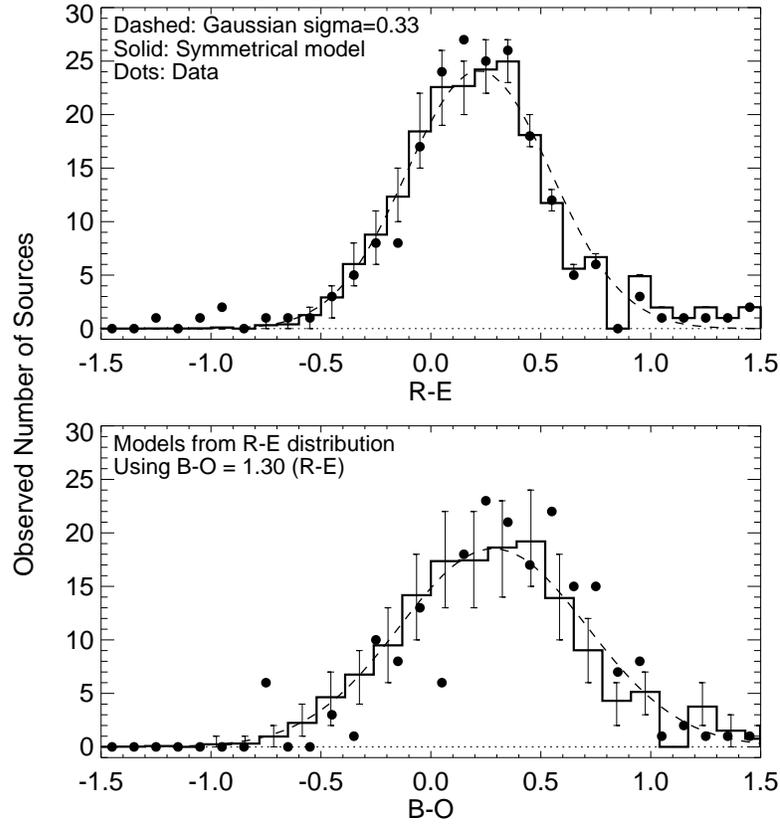}
\caption{
Observed distribution of magnitude change $\delta m = m_{CCD}-m_{POSS}$
for quasars in our sample.  Objects on the right were fainter in the
current CCD observations.  The histogram is skewed due to Malmquist
bias (see text for discussion.) The dots are the observed counts, and
the solid line shows the predicted counts in bins (with $1\sigma$
uncertainty error bars) from a symmetrical, Malmquist-corrected model.
The dashed line is a Gaussian distribution having the same rms as the
model distribution.  (a) Red magnitude differences ($m_R-m_E$). The
Gaussian width is $\sigma=0.33$. Error bars account for the use of these
same data to compute the model values and show the expected range for data
points about that mean. (b) Blue magnitude differences ($m_B-m_O$). The
model (both histogram and Gaussian) are derived from the red
delta-magnitude distribution using the relation $B-O = 1.3(R-E)$.  In
this case the model is derived completely independently from the data,
so the error bars show the expected range for these data about the mean
due to Poisson statistics.
}
\label{figobsdist}
\end{figure*}
}

Figure~\ref{figobsdist} shows the distributions of the change in quasar
magnitude ($\delta m = m_{CCD}-m_{POSS}$) in the red and blue bands.
The distributions are clearly skewed, with many more quasars found to
be dimmer in the CCD data than are found to be brighter.  This
asymmetry is the result of Malmquist bias (Malmquist 1924) in the FBQS
sample.  Because the quasar number counts increase steeply with
magnitude, a disproportionate fraction of quasars near the FBQS
magnitude limit are included in the sample because small brightness
fluctuations at the time of the POSS-I observation moved them above the
optical threshold for inclusion in the sample.

It is relatively straightforward to correct the observed $\delta m$
distribution, $P_{obs}(\delta m)$, for Malmquist bias and thus
determine the true distribution, $P_{true}(\delta m)$.  Let $m$ be the
true (current-epoch CCD) magnitude of the quasar, $m_0$ be the
magnitude brightness limit of the FBQS sample, and $N(m)$ be the
cumulative number of quasars brighter than $m$.  For $\delta m>0$, the
quasars were brighter on the POSS-I plates than they are currently, so
the FBQS sample was effectively looking deeper in the $N(m)$
distribution and thus included more objects.  The observed magnitude
distribution is
\begin{equation}
P_{obs}(\delta m) = {P_{true}(\delta m) N(m_0+\delta m) \over
  \int_{-\infty}^{\infty} P_{true}(\delta m) N(m_0+\delta m) d\delta m}
  \quad .
\end{equation}
The factor in the denominator simply normalizes $P_{obs}$ so that
its integral is unity.

For quasars at $V \sim 17.5$, the cumulative number distribution is
approximately exponential:
\begin{equation}
N(m) = N_0 e^{(m-m_0) \ln F} \quad ,
\end{equation}
where $F$ is the factor by which the counts increase for a 1 magnitude
change ($F\simeq7.6$, La Franca and Cristiani 1997).  Thus the observed
distribution is
\begin{equation}
P_{obs}(\delta m) = {P_{true}(\delta m) e^{\delta m \ln F}\over
  \int_{-\infty}^{\infty} P_{true}(\delta m) e^{\delta m \ln F} d\delta m}
  \quad .
\end{equation}
This integral can be evaluated analytically for a Gaussian distribution.
If $P_{true} = \exp(-{\delta m}^2/2\sigma^2)$ then
\begin{equation}
P_{obs}(\delta m) = \exp(-[\delta m-\sigma^2\ln F]^2/2\sigma^2)
\quad .
\end{equation}
A Gaussian distribution simply has its mean shifted by
$\sigma^2\ln F = 2.03\sigma^2$ for $F=7.6$.

More generally, it is safe to assume that the true distribution of
$\delta m$ is symmetrical about zero.  Using this (weak) assumption, we
can symmetrize the observed histogram of $\delta m$.  Consider two
histogram bins symmetrically placed about zero:
$|\delta m - \delta m_h| < b$ and $|\delta m + \delta m_h| < b$, where
the bins are centered at $\pm \delta m_h$ and the bin width is $2b$.
Let the number of objects in each $\delta m$ bin be $N_+$ and $N_-$.
By the symmetry assumption, the true distribution has the same expected
number of objects in each bin, $\bar N$.  If the bin width is small,
the expected number of objects $E(N)$ in the observed distribution is
modified by a Malmquist bias factor $W$:
\begin{eqnarray}
E(N_+) & = & {\bar N} W_+ = {\bar N} \exp(\delta m_h \ln F)
	\quad \hbox{, and} \\
E(N_-) & = & {\bar N} W_- = {\bar N} \exp(-\delta m_h \ln F)
	\quad .
\end{eqnarray}
The most accurate estimator of the true number of objects in the bin is
then
\begin{equation}
{\bar N} = {N_+ + N_- \over W_+ + W_-} \quad .
\end{equation}
Some straightforward algebra will convince the reader that this gives
the correct answer in the case of a Gaussian distribution (except for a
constant normalization factor, which can be computed directly from
${\bar N}$.) The uncertainty in the estimator ${\bar N}$ is easily
determined, since it has a Poisson noise distribution scaled by the
$W_+ + W_-$ factor.

A subtle point is that the observed blue magnitude differences
are biased only through their coupling to the red magnitudes, because
the red magnitude was used to define the sample.  There is a weak
bias that directly affects the blue due to a limit $O-E<2$ on the
colors of FBQS quasars.  Thus, quasars that are near the FBQS limit
$E=17.8$ must be brighter than $O=19.8$ to be included in the sample.
However, most quasars are much bluer than this limit and so the
bias in the blue $\delta m_B$ distribution (which is apparent in
Fig.~\ref{figobsdist}) appears only because the blue variation is
highly correlated with the red variation.

To understand how the bias affects the blue $\delta m_B$
distribution, consider a hypothetical population of objects
that do not vary at all in the red but do vary in the blue.
For simplicity we also assume measurement errors are zero.
At the FBQS discovery epoch, we simply select all the objects
brighter than the sample magnitude limit.  Some of those objects
are brighter than average in the blue and some are fainter, but
all are included in the sample because there is no discrimination
against faint blue magnitudes.  At the current epoch we would
find that the blue magnitudes differ but the red magnitudes are
unchanged.  The observed blue $\delta m_B$ distribution would
consquently be symmetrical about zero, with no Malmquist bias at all.

Now consider another extreme hypothetical population, where both the
red and blue magnitudes of the objects vary but the blue magnitude
is completely determined by the red magnitude, such that
$\delta m_B = \alpha\,\delta m_R$ where $\alpha$ is a constant.
In this case, the selected sample is biased by brightness fluctuations
that bring objects with brighter $R$ magnitudes into the sample; those
objects also have brighter $B$ magnitudes.  At the current epoch,
we therefore find that many objects have dimmed in both $B$ and $R$
according to the hypothesized linear relationship.  In this case the
observed $\delta m_R$ distribution is related to the true distribution
as derived in equations~(1)--(7) above, and the $\delta m_B$ distribution
is merely a rescaled version of the $\delta m_R$ distribution.  For
example, if $\alpha=2$, then the $\delta m_B$ distribution will be
twice as wide, shifted twice as far from zero, and (to conserve counts)
half as high as the $\delta m_R$ distribution.  This result is
{\it not} the same as if the blue magnitudes were themselves Malmquist-biased;
in that case, when the distribution is twice as wide, it is shifted
four times as far from zero (Eq.~4) rather than twice as far.

\placefigure{figredvsblue}

\preprint{
\begin{figure*}
\centering \leavevmode
\epsfxsize=0.6\textwidth \epsfbox{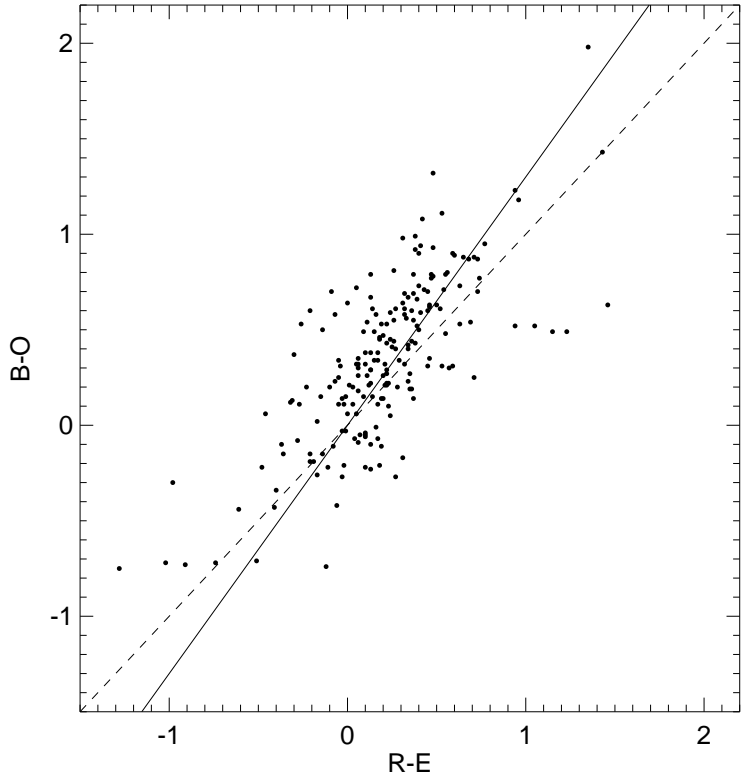}
\caption{
A comparison of the $R-E$ and $B-O$ variations for our quasar sample.
The dashed line indicates identical variations in the two colors, while
the solid line has the best fit slope of 1.30.  The most highly
variable source ($\delta m_R \sim 4.5,\delta m_B \sim 4.2$) falls
outside the plot. The magnitudes are strongly correlated, with higher
variability amplitudes in the blue.
}
\label{figredvsblue}
\end{figure*}
}

The real case is, of course, between these two extremes, with the blue
variation having components that are both correlated with and
independent of the red variation.  Fig.~\ref{figredvsblue} shows that
$\delta m_B$ is in fact strongly correlated with $\delta m_R$.  The
best-fit line through the points, allowing for measurement errors in
both coordinates, is $\delta m_B = 1.30\, \delta m_R$.  If we assume that
the uncertainty in $\delta m_R = 0.17$ (an estimate of the combined
measurement errors in $E$ and $R$), then the scatter about the best fit
line indicates that the uncertainty in $\delta m_B \simeq 0.24$ in
order that the $\chi$-square of the fit be approximately 1 per degree
of freedom.  Since the measurement errors on $B$ and $O$ are similar to
those in the red, the implied ``intrinsic'' scatter in
$\delta m_B \simeq 0.17$.  In other words, the portion of the blue variability
that is uncorrelated with the red variability is relatively modest and is
of the same order as the measurement error.

\placefigure{figcorrdist}

\preprint{
\begin{figure*}
\centering \leavevmode
\epsfxsize=0.6\textwidth \epsfbox{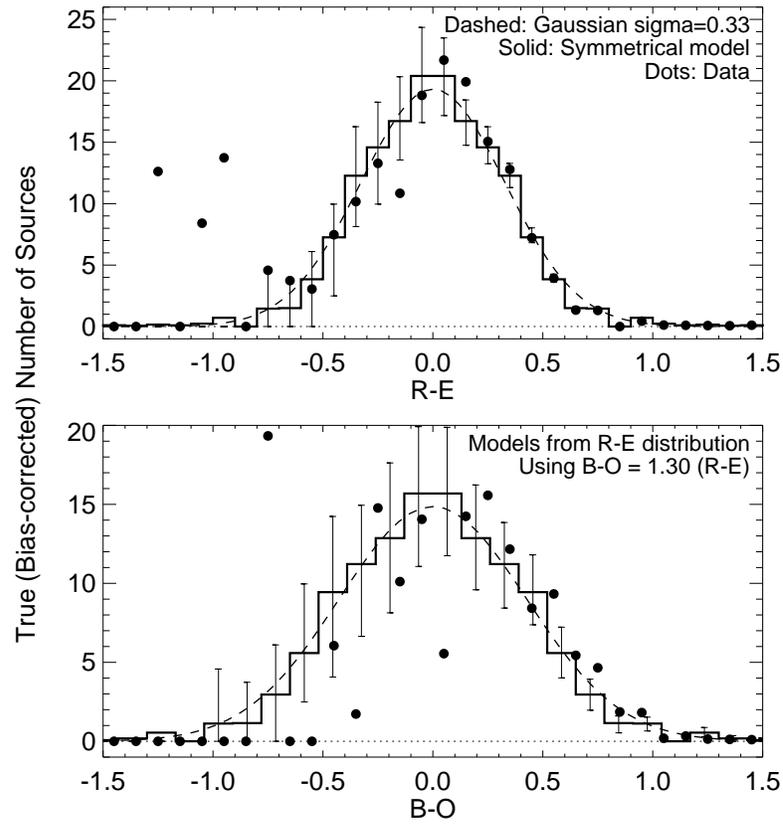}
\caption{
The distribution of magnitude differences corrected for Malmquist bias
under the assumption that the true distribution is symmetrical about
zero.  As in Fig.~2, the dots are the observed counts corrected for the
Malmquist factor in each bin, the solid line is the true distribution
of $\delta m$, and the dashed line is a model Gaussian; in both panels,
the models are derived from the red data.  (a) Red magnitude
differences. (b) Blue magnitude differences.
}
\label{figcorrdist}
\end{figure*}
}

Figure~\ref{figcorrdist} shows the $\delta m$ distributions for both
the red and blue symmetrized and corrected for the Malmquist bias (see
Eq.~7).  The predictions from the symmetrical model are in good
agreement with the data.  Note in particular the reasonably good
agreement with the blue distribution; since the model in this case was
derived from the red $\delta m_R$ data, there are no free parameters in
the fit.  The red distribution was converted to a blue model simply by
scaling the $\delta m$ values by a factor 1.30, as indicated by the fit
in Fig.~\ref{figredvsblue}.  A Gaussian with the same rms as the model
distribution is also shown and appears to be quite a good
representation for the bulk of the distribution, although there is
certainly a non-Gaussian tail of high-amplitude variables.

The predictions of the symmetrical model (and the associated Gaussian)
are also shown in Fig.~\ref{figobsdist} where they are overlain on the
observed (biased) counts.  This predicted distribution is obtained by
multiplying the true symmetrical distribution by the Malmquist factor
$W$ from Eq.~(4).  As was mentioned above, the biased Gaussian
distribution is just a shifted Gaussian with the same width.  Note that
the Gaussian in Fig.~\ref{figobsdist}a has only {\it one} free
parameter (the width) because the shift is {\it specified} by the width
and the normalization is determined by the total number of quasars;
thus the good agreement with the observational distribution supports
both the general approach adopted here and the idea that there is a
roughly Gaussian underlying distribution of amplitude fluctuations.

In the following section we examine the dependence of variability on
various quasar properties (e.g., redshift and radio-loudness).
Malmquist bias affects all of these measurements.  The correction
procedure derived above can also be applied to any subset of the data
to determine a Malmquist-corrected rms value for the variability of
that subset.  The approach we have adopted is to compute the
symmetrized distribution for the subset (typically objects having some
property falling within a moderately narrow range) and then calculate
the rms from that distribution. From the analysis of the whole sample
above, this approach is seen to give a good measurement of the width of
the (approximately Gaussian) histogram.  We exclude the single object
that varied by 4 magnitudes (4C38.41) from these calculations since it
produces a wild outlying value for the rms of any bin into which it
falls.

\subsection {Correlation with other quasar properties}

Figure~\ref{figredvsblue} compares the variability amplitudes for the
red and blue bands. The correlation between the two bands is apparent,
as is the fact that the average variability amplitude is higher in the
blue. This trend has been seen in previous optically-selected samples;
e.g., Figure~2 in Trevese et al. (1997) suggests that for a factor of
two change in wavelength, $\delta m_B -\delta m_R \sim 0.11$. For our
sample, the mean difference between the variability amplitudes is 0.13
mag.  When the Malmquist bias is removed, the rms variations in the
blue and red are 0.45 mag and 0.33 mag, respectively
(Fig.~\ref{figobsdist}).  If we subtract in quadrature the 0.17 mag rms
contributed by calibration errors (\S4), the quasar blue and red rms
values are 0.42 and 0.28 magnitudes.  Thus the bias-corrected rms
difference is 0.14 mag, in good agreement with Trevese et al.

In Figure~\ref{figredshift}, we show the distribution of $\delta m$ as
a function of redshift.  We have made no correction for emission lines
in our analysis; in that $B,O$ and $R,E$ are not identical bands,
strong lines entering the bands at slightly different redshifts could
be a cause for concern. We have marked the locations at which MgII,
CIV, and Ly$\alpha$ enter and leave the POSS-I bands; no significant
anomalies are seen. While it is possible that a few objects with very
strong lines could have their $\delta m$ values affected by line
emission, it does not appear that the statistical conclusions derived
from our sample will be biased by this effect. Note again the strong
preference for positive $\delta m$ values induced by the Malmquist
bias.  However, the amplitude of the variability in both bands is
independent of $z$.  Binning the objects in redshift intervals of 0.5,
we find, for example, that $<\delta m_R>$ differs by only 0.01 mag in
the $z<0.5$ and $z>3$ bins, and in none of the seven bins is the mean
value more than $1.5\sigma$ from the overall mean value of 0.33 when
correction for the Malmquist bias is applied (see above). These results
are inconsistent with the majority of recent studies on optically
selected samples (e.g., Fernandes, Aretxaga, and Terlevich 1996 and
references therein). Given the complex interaction of k-corrections,
wavelength-dependent variability, the redshift-luminosity correlation
in a flux-limited sample, etc., it is difficult to assess the
significance of this disagreement.

The range of redshifts in the sample (and the range of time over which
the data at the two epochs were collected) means the proper time
intervals sampled range from 10 to 45 years. The mean values of the
variability amplitudes for proper times less than, and greater than, 20
years are identical to within 0.02 and 0.01 magnitudes, the respective
errors in the means for the blue and red bands.

\placefigure{figredshift}

\preprint{
\begin{figure*}
\centering \leavevmode
\epsfxsize=0.6\textwidth \epsfbox{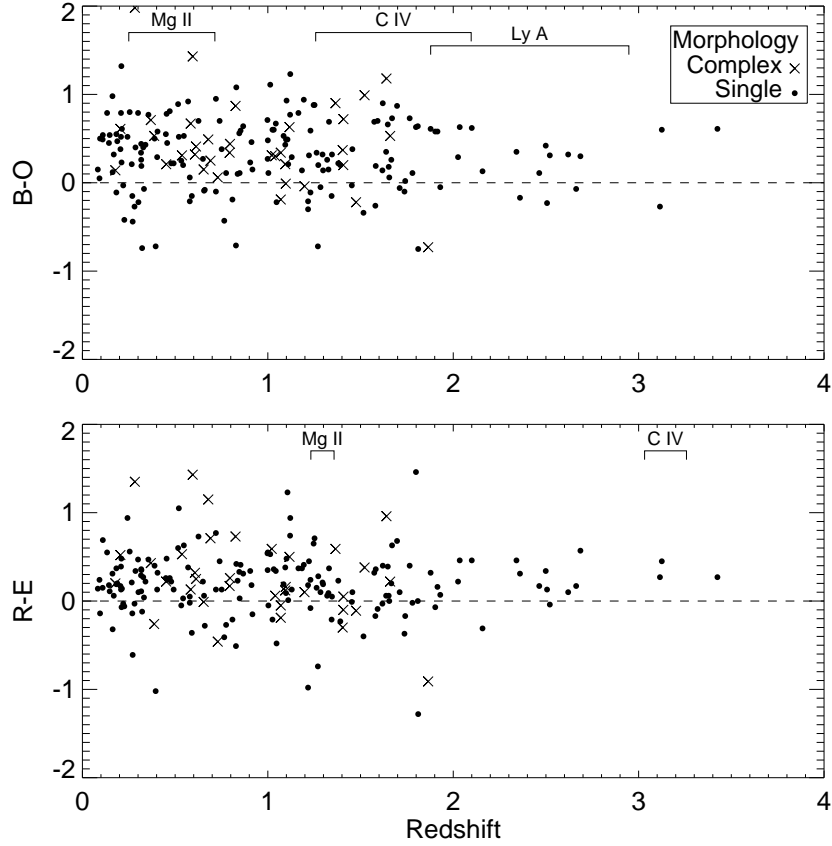}
\caption{
A plot of $\delta m_R$ and $\delta m_B$ as a function of redshift.
Complex (FR~II and core-jet) extended radio sources are indicated by
$\times$ symbols; simple single sources are indicated by dots. The
redshifts where various strong lines enter and leave the POSS-I bands
are indicated at the top of each panel.  While a strong bias towards
sources becoming fainter is apparent in both bands, there is no obvious
trend in variability amplitude with redshift, contrary to the results
from some recent optically selected samples.
}
\label{figredshift}
\end{figure*}
}

Figures~\ref{figradiolum}a and b illustrate the size of the effect
Malmquist bias can introduce when searching for correlations between
variability amplitude and other quasar properties. We show the median
amplitudes over a span of $10^7$ in radio luminosity both before and
after correcting for the Malmquist bias using the prescription outlined
above. While before correction, the plots are noisy but suggest a
positive correlation with luminosity in the R band, the corrected
values are flat, with all points consistent with the median value to
within less than $2\sigma$. We also find that variability amplitude is
uncorrelated with optical apparent magnitude when Malmquist-corrected
distributions are employed.

The high resolution of the {\it FIRST} survey allows us to comment on
the radio morphology dependence of the variability. The FBQS selection
criteria require a radio component coincident with the optical
counterpart, so our survey may be partially incomplete for sources with very weak
core components and dominant radio lobes (classical FR-II radio
doubles), although our recent extension of the FBQS to include candidate
doubles suggests the incompleteness
is less than 5\% (Becker et al. 2001). Furthermore, it is possible that
the lobe components for some of our identified quasars have been resolved
out in our high-resolution radio observations, although, again, the fraction of
such objects is small, since most have not been found to have extended components in the much lower-resolution NVSS survey (Condon et al. 1998). 

In total, roughly 10\% of the sources in the current
sample show extended emission, and most of these are FR-II objects. We
have computed the rms($B-O$) and rms($R-E$) values for the single and
complex sources separately and, surprisingly, find that those sources
with extended compnents are slightly {\it more} variable; e.g.,
rms($R-E) = 0.320 \pm 0.010$ for the single sources, while rms($R-E) =
0.376 \pm 0.028$ for the complex objects. This result is not highly
significant ($\sim 2\sigma$) but is in the opposite sense expected from
standard unified models of AGN in which the sources with extended lobes
are oriented roughly normal to the line of sight, while sources with
bright core components have their jets beamed toward us and should
thus, on average, be more variable. Given possible incompleteness in the
sample and the low significance of the distinction, it is clear
that a larger sample of FR-II sources would be required to draw quantitative 
conclusions.

\placefigure{figradiolum}

\preprint{
\begin{figure*}
\begin{tabular}{cc}
\epsfxsize=0.45\textwidth \epsfbox{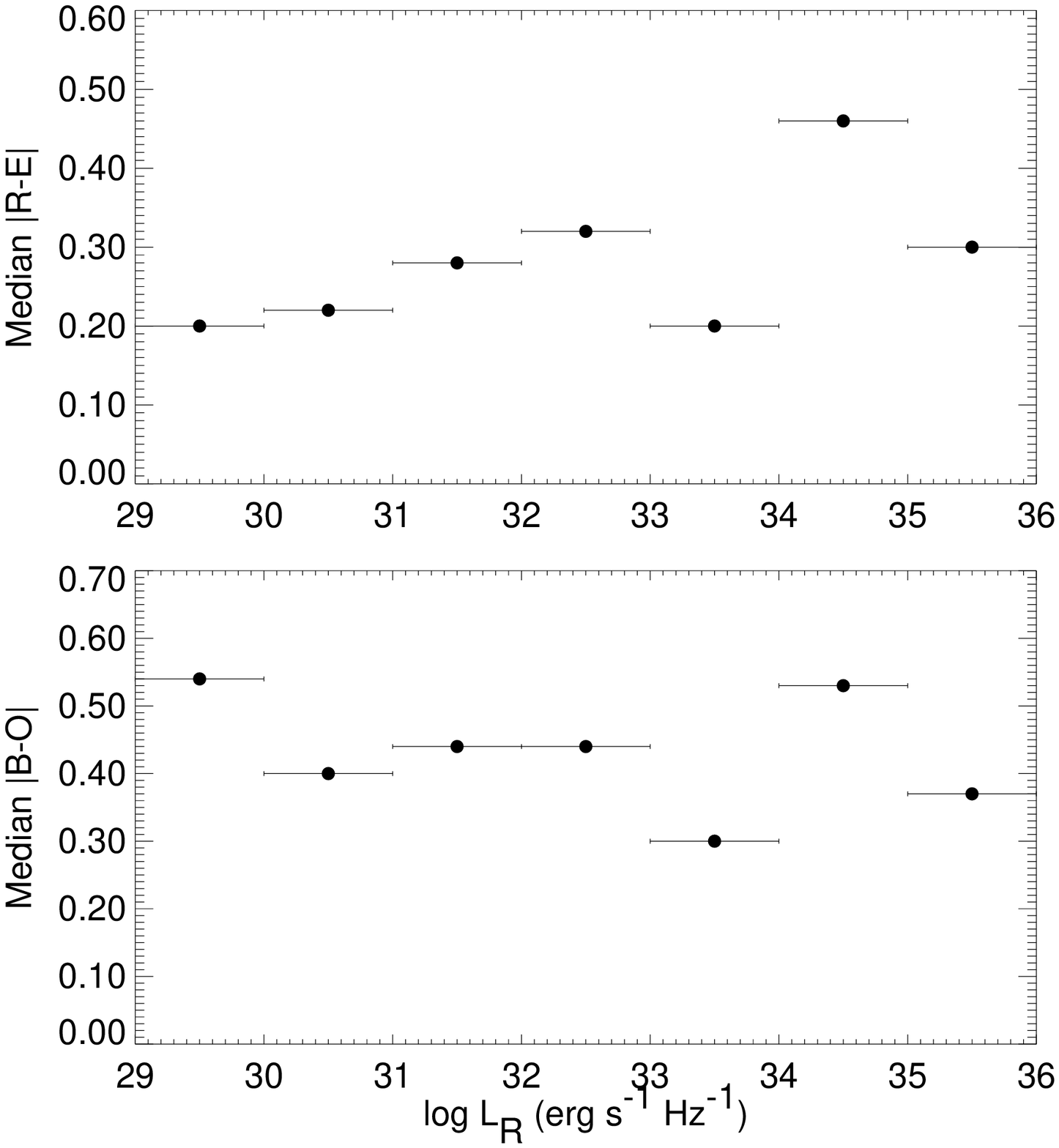} &
\epsfxsize=0.45\textwidth \epsfbox{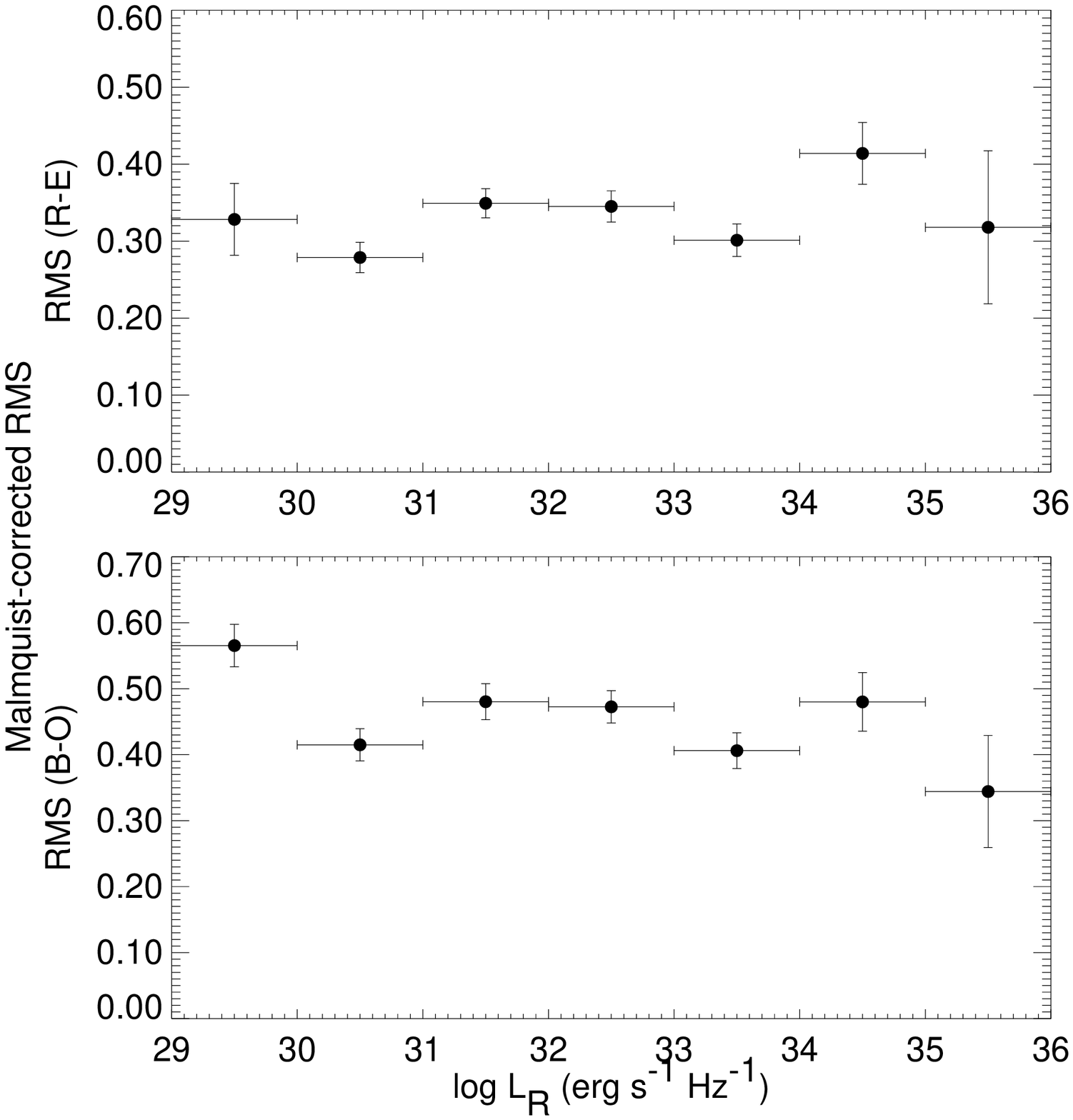} \\
(a) & (b) \\
\end{tabular}
\caption{
The distribution of median $\delta m_R$ and $\delta m_B$ as a function
of radio luminosity. Fig.~6a shows this distribution before correction
for Malmquist bias, while 6b shows the result of applying the
correction stipulated in \S5.2. After the correction, no dependence on
$L_r$ is apparent.
}
\label{figradiolum}
\end{figure*}
}

Fig.~\ref{figlogr} shows the dependence of variability amplitude on the
radio-loudness parameter over a range in $R^*$ of 10,000. The curves in
the two bands are remarkably similar, and suggest a nearly constant
amplitude over most of the range, with a possible turn up for the
extreme radio-loud objects.

\placefigure{figlogr}

\preprint{
\begin{figure*}
\centering \leavevmode
\epsfxsize=0.6\textwidth \epsfbox{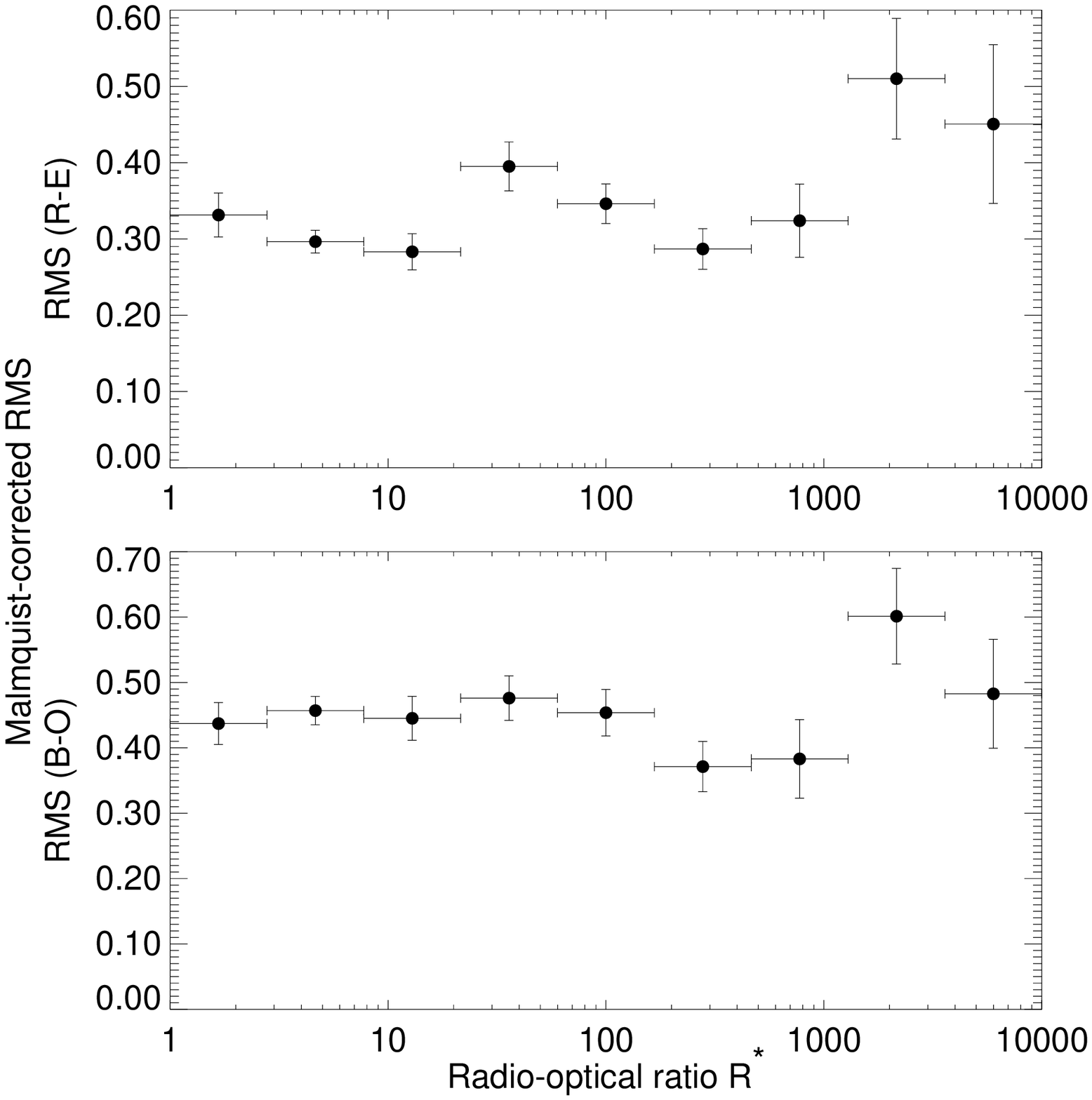}
\caption{
Variability amplitude is displayed as a function of the radio loudness
parameter $R^*$ (using the Malmquist-corrected rms values of $\delta m$).
In both bands, the distributions are flat, with the hint of an upturn
for the radio-loudest objects.
}
\label{figlogr}
\end{figure*}
}

The final figure (Fig.~\ref{fig3d}) shows a three-dimensional
distribution of variability amplitude as a function of both $R^*$ and
absolute magnitude. Fluctuations from the mean value of less than
$1\sigma$ have been suppressed (by shrinking all values toward the
mean) to emphasize the more significant features of the distribution.
The trend noted by Hook et al. (1994) in their optically selected
sample for variability amplitude to decrease with increasing absolute
magnitude is apparent here for the objects with $R^*<10$ (the
radio-quiet portion of our sample)\footnote{Comparing the radio quiet
sources with $M_B<-28$ to those in the range $-28<M_B<-24$, the rms
variability amplitudes are different at the $6.1 \sigma$ level.}, but
disappears for objects with higher radio-to-optical flux ratios. The
dependence of this trend on radio loudness could arise if the overall
variability is composed of two parts: one due to long-term changes in
the structure of the accretion disk giving rise to changes in the
thermal tail of the `big blue bump', and the other arising from
fluctuations in the non-thermal, beamed emission. The most luminous
objects with large, slowly varying disks would have a smaller amplitude
of variability unless the nonthermal emission comes to dominate the
optical flux; the higher amplitude variability for the radio-loudest
objects, apparent in the figure, is consistent with assigning the
greater variability amplitude to the beamed emission. 

The decrease in
variability seen here for the lowest luminosity objects is significant
at the $4.9\sigma$ level (comparing $M_B>-24$ to $-24>M_B>-28$) and has
not been reported previously as a consequence of the exclusion from
most quasar samples of objects with $M_B > -23.5$. Furthermore, the
trend apparently does not continue toward lower luminosity Seyfert
galaxies; in the compilation by Winkler et al. (1992), six of the seven
highest amplitude variables over a four-year period are drawn from the
upper half of the luminosity distribution (median luminosity
$M_B=-21.2$).

\placefigure{fig3d}

\preprint{
\begin{figure*}
\centering \leavevmode
\epsfxsize=0.9\textwidth \epsfbox{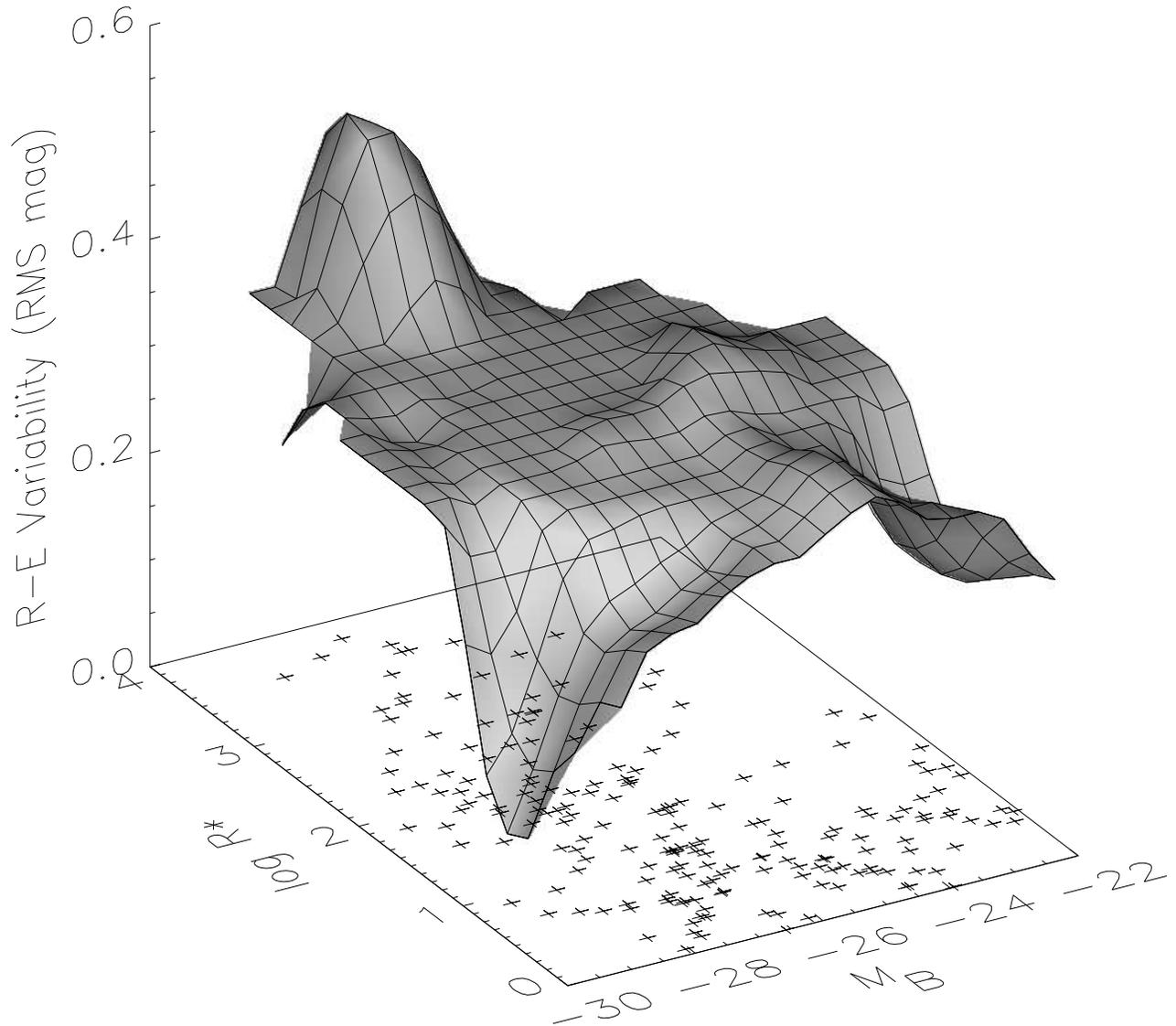}
\caption{
Red variability amplitude is displayed as a function of the radio
loudness parameter $R^*$ and the absolute blue magnitude. The
distribution of the points on the $\log R^*$ - $M_B$ plane is
indicated. The surface has been plotted such that all variations from
the mean with a significance of less than $1 \sigma$ are suppressed (by
shrinking the values toward the mean value, 0.34) to emphasize the more
significant features of the distribution.
}
\label{fig3d}
\end{figure*}
}

\section {Summary and conclusions}

We have presented a study characterizing the long-term ($\sim45$ yr)
variability properties for a radio-selected sample of 201 quasars
spanning large ranges in redshift, luminosity, and radio loudness. We
find little or no dependence of the variability amplitude on radio or
optical flux, radio luminosity, or redshift. We do see higher mean
amplitudes in the blue band than the red band, consistent with the
reported wavelength dependence in optical samples. The dependence on
radio loudness is largely flat, with a slight upturn for the extremely
radio-loud objects. The decrease in variability amplitude for optically
luminous quasars is confined to radio-quiet objects.  We also
demonstrate the importance of considering Malmquist bias when
characterizing the variability properties of a quasar sample, and
describe a technique for correcting observed distributions of
variability amplitude.

The complex interdependence of the observed variability on frequency,
redshift, bolometric luminosity, radio loudness, and possibly other
parameters makes it difficult to draw definitive conclusions regarding
the physics of the central engine or quasar evolution from a study such
as this. A larger study of a carefully selected sample of FBQS objects,
supplemented by a sample of much fainter {\it FIRST} counterparts,
could provide the breadth of coverage in luminosity over a broad range
of redshifts necessary to begin to untangle these effects.

\acknowledgments

The success of the {\it FIRST} survey is in large measure due to the
generous support of a number of organizations. In particular, we
acknowledge support from the NRAO, the NSF (grants AST-98-02791 and
AST-98-02732), the Institute of Geophysics and Planetary Physics
(operated under the auspices of the U.S. Department of Energy by
Lawrence Livermore National Laboratory under contract No.
W-7405-Eng-48), the Space Telescope Science Institute, the National
Geographic Society (grant NGS No.~5393-094), and Columbia University.
DJH is grateful for the support of the Raymond and Beverly Sackler
Fund, and joins RHB and RLW in thanking the Institute of Astronomy at
the University of Cambridge for hospitality during much of this work.
We would also like to thank the Editor Paul Hodge and the referee Paul Wiita
for stunning speed and efficiency in taking this paper from submission
through refereeing, revision, and resubmission in four days.
This paper is Contribution Number 689 of the Columbia Astrophysics
Laboratory.

{}

\begingroup
\tabcolsep=3pt
\footnotesize
\begin{deluxetable}{ccccrcrccrrrrccrl}
\tablewidth{0pt}
\tablecaption{FIRST Variable Quasar Catalog}
\tablenum{2}
\tablehead{
\colhead{RA} & \colhead{Dec} & \colhead{POSS1 obs} & \colhead{CCD obs} & \colhead{B} & \colhead{O} & \colhead{B-O} & \colhead{R} & \colhead{E} & \colhead{R-E} & \colhead{$S_p$} & \colhead{$S_i$} & \colhead{z} & \colhead{M(B)} & \colhead{L(R)} & \colhead{$\log R^*$} & \colhead{Morphology}\\
\colhead{(1)}&\colhead{(2)}&\colhead{(3)}&\colhead{(4)}&\colhead{(5)}&\colhead{(6)}&\colhead{(7)}&\colhead{(8)}&\colhead{(9)}&\colhead{(10)}&\colhead{(11)}&\colhead{(12)}&\colhead{(13)}&\colhead{(14)}&\colhead{(15)}&\colhead{(16)}&\colhead{(17)}
}
\startdata
07 17 34.48&+29 16 13.7& 1953.197& 1996.277&  18.71& 17.78&  0.93&  17.86& 17.38&  0.48&     2.18&     2.25& 1.101& -26.7&  31.8&  0.65&            \nl
07 25 26.45&+29 13 40.7& 1953.197& 1996.277&  18.45& 18.13&  0.32&  17.44& 17.39&  0.05&     1.64&     0.97& 1.346& -26.8&  31.8&  0.63&            \nl
07 25 50.59&+28 19 06.1& 1953.197& 1997.268&  18.18& 18.25& -0.07&  17.37& 17.20&  0.17&    37.99&    38.85& 2.663& -28.4&  33.8&  1.95&            \nl
07 29 52.34&+30 46 44.1& 1953.197& 1996.277&  18.26& 17.72&  0.54&  17.22& 17.11&  0.11&     1.48&     1.17& 0.147& -22.1&  29.9&  0.57&            \nl
07 44 12.08&+29 59 06.9& 1953.197& 1996.277&  18.93& 18.06&  0.87&  18.59& 17.91&  0.68&    93.72&    94.90& 1.697& -27.5&  33.8&  2.33&            \nl
07 46 04.95&+29 22 50.4& 1955.279& 1996.277&  18.32& 17.79&  0.53&  17.92& 17.29&  0.63&     1.62&     1.31& 0.547& -25.0&  31.0&  0.57&            \nl
07 48 04.65&+37 40 45.7& 1955.192& 1996.279&  19.16& 18.55&  0.61&  18.37& 18.05&  0.32&     2.50&     2.16& 1.878& -27.2&  32.3&  0.93&            \nl
07 49 48.24&+34 54 44.2& 1955.192& 1996.277&  17.10& 16.31&  0.79&  16.46& 15.91&  0.55&     1.31&     0.82& 0.133& -23.3&  29.7& -0.04&            \nl
07 50 47.34&+41 30 33.0& 1953.279& 1996.279&  17.59& 17.45&  0.14&  17.03& 16.66&  0.37&     2.06&     2.22& 1.184& -27.2&  31.8&  0.50&            \nl
07 51 12.34&+29 19 38.2& 1955.279& 1996.277&  16.14& 15.68&  0.46&  15.63& 15.45&  0.18&     1.00&     1.16& 0.912& -28.3&  31.3& -0.46&            \nl
07 54 48.89&+30 33 54.9& 1955.279& 1996.277&  18.07& 17.63&  0.44&  17.44& 17.18&  0.26&    44.75&    46.47& 0.796& -26.1&  32.8&  1.94&core-jet 12s\nl
07 54 58.35&+29 41 54.1& 1955.279& 1996.277&  18.67& 18.05&  0.62&  18.29& 17.83&  0.46&   398.19&   410.00& 2.100& -28.0&  34.6&  2.93&            \nl
07 58 00.12&+39 20 29.8& 1955.159& 1996.279&  14.83& 14.33&  0.50&  14.09& 14.23& -0.14&    10.80&    11.56& 0.095& -24.5&  30.4&  0.12&            \nl
07 59 28.30&+30 10 28.1& 1955.279& 1996.279&  18.09& 17.82&  0.27&  17.60& 17.25&  0.35&   161.38&   182.20& 0.999& -26.4&  33.6&  2.58&            \nl
08 02 20.51&+30 35 42.8& 1955.279& 1996.279&  19.42& 18.24&  1.18&  18.85& 17.89&  0.96&    27.81&    28.55& 1.640& -27.2&  33.2&  1.88&FRII 55s    \nl
08 04 09.30&+38 53 49.1& 1955.159& 1996.279&  18.20& 17.97&  0.23&  16.92& 16.99& -0.07&     2.88&     2.68& 0.211& -22.6&  30.5&  0.95&            \nl
08 08 49.25&+40 09 18.7& 1955.159& 1996.279&  18.71& 18.12&  0.59&  18.01& 17.60&  0.41&     1.95&     1.49& 0.853& -25.7&  31.5&  0.75&            \nl
08 09 06.24&+29 12 35.3& 1955.279& 1996.279&  17.43& 17.65& -0.22&  17.01& 17.12& -0.11&    21.48&    21.87& 1.476& -27.5&  33.0&  1.55&FRII 140s   \nl
08 09 06.52&+41 39 32.5& 1955.159& 1996.279&  17.70& 17.39&  0.31&  17.12& 16.67&  0.45&     2.21&     2.45& 1.222& -27.3&  31.9&  0.51&            \nl
08 11 30.81&+39 04 11.2& 1955.159& 1996.279&  18.71& 18.05&  0.66&  17.92& 17.53&  0.39&     1.62&     1.82& 1.647& -27.4&  32.0&  0.61&            \nl
08 11 48.30&+37 56 45.6& 1954.227& 1996.279&  18.14& 17.76&  0.38&  17.24& 17.14&  0.10&     1.76&     1.60& 1.456& -27.4&  31.9&  0.50&            \nl
08 14 15.07&+41 23 23.4& 1955.159& 1996.279&  18.55& 18.23&  0.32&  17.83& 17.62&  0.21&     5.15&     4.65& 1.295& -26.6&  32.3&  1.17&            \nl
08 21 07.63&+31 07 51.1& 1955.301& 1996.279&  17.44& 17.12&  0.32&  16.76& 16.66&  0.10&    85.41&    89.98& 2.620& -29.5&  34.1&  1.87&            \nl
08 23 58.21&+36 49 53.8& 1953.107& 1996.279&  19.31& 18.41&  0.90&  18.54& 17.95&  0.59&    13.64&    13.64& 1.365& -26.6&  32.7&  1.66&core-jet 20s\nl
08 24 06.26&+33 42 45.3& 1953.107& 1997.184&  18.46& 18.20&  0.26&  17.85& 17.74&  0.11&     1.93&     1.08& 0.317& -23.3&  30.6&  0.85&            \nl
08 24 55.54&+39 16 42.4& 1955.159& 1996.282&  17.71& 18.01& -0.30&  16.87& 17.85& -0.98&  1403.96&  1456.09& 1.217& -26.7&  34.7&  3.54&            \nl
08 32 25.35&+37 07 36.1& 1953.107& 1996.282&  16.66& 16.61&  0.05&  16.02& 15.78&  0.24&    11.78&    11.73& 0.092& -22.1&  30.4&  1.04&            \nl
08 32 46.94&+28 53 11.8& 1955.301& 1996.281&  17.54& 17.96& -0.42&  17.02& 17.08& -0.06&     1.64&     1.73& 0.226& -22.8&  30.3&  0.72&            \nl
08 36 36.90&+41 25 54.6& 1953.285& 1996.282&  18.11& 17.97&  0.14&  17.60& 17.41&  0.19&   424.23&   443.13& 1.298& -26.9&  34.2&  3.00&            \nl
08 40 44.40&+36 33 27.6& 1953.107& 1996.282&  17.41& 17.52& -0.11&  16.42& 16.50& -0.08&     1.63&     1.00& 1.230& -27.2&  31.7&  0.39&            \nl
08 41 18.08&+35 44 38.4& 1953.107& 1996.282&  17.02& 16.91&  0.11&  16.46& 16.48& -0.02&     6.01&     5.66& 1.780& -28.7&  32.6&  0.66&            \nl
08 44 37.90&+41 24 31.5& 1953.285& 1996.282&  17.74& 17.22&  0.52&  17.50& 16.45&  1.05&     1.58&     1.38& 0.520& -25.5&  31.0&  0.34&            \nl
08 47 16.08&+37 32 17.0& 1953.107& 1996.282&  18.22& 17.44&  0.78&  17.92& 17.44&  0.48&     1.69&     1.20& 0.454& -24.9&  30.9&  0.47&            \nl
08 49 02.57&+30 02 35.0& 1954.022& 1996.282&  16.59& 16.67& -0.08&  16.37& 16.65& -0.28&     1.02&     1.26& 0.660& -26.6&  31.1&  0.00&            \nl
08 55 08.94&+34 14 41.2& 1955.277& 1996.282&  19.03& 18.32&  0.71&  18.13& 17.59&  0.54&     1.21&     1.25& 0.998& -25.9&  31.4&  0.62&            \nl
08 57 21.07&+36 38 44.3& 1955.277& 1996.282&  18.26& 17.71&  0.55&  17.91& 17.65&  0.26&     1.09&     0.78& 0.450& -24.6&  30.7&  0.38&            \nl
09 00 08.02&+36 46 10.4& 1955.277& 1996.282&  18.11& 17.22&  0.89&  17.70& 17.10&  0.60&     2.67&     2.38& 0.516& -25.4&  31.2&  0.57&            \nl
09 09 47.85&+31 24 43.2& 1955.310& 1996.282&  18.10& 17.89&  0.21&  16.89& 16.68&  0.21&     1.81&     1.50& 0.266& -23.2&  30.5&  0.71&            \nl
09 10 54.16&+37 59 14.9& 1955.277& 1996.282&  17.43& 17.30&  0.13&  17.15& 17.46& -0.31&   250.93&   265.93& 2.158& -28.8&  34.4&  2.44&            \nl
09 12 06.95&+39 11 45.7& 1953.364& 1996.282&  18.48& 18.58& -0.10&  17.77& 18.14& -0.37&    16.98&    17.80& 1.737& -27.0&  33.1&  1.81&            \nl
09 12 47.80&+28 54 06.0& 1955.310& 1996.282&  18.33& 17.78&  0.55&  17.16& 16.79&  0.37&     1.22&     0.88& 0.183& -22.5&  30.0&  0.51&            \nl
09 13 28.21&+39 44 43.8& 1953.364& 1996.282&  18.25& 18.51& -0.26&  17.38& 17.55& -0.17&     2.06&     2.09& 1.580& -26.8&  32.1&  0.86&            \nl
09 19 54.30&+29 14 08.5& 1955.310& 1996.279&  17.86& 17.96& -0.10&  17.61& 17.48&  0.13&     8.10&     8.33& 0.720& -25.5&  32.0&  1.33&            \nl
09 23 35.46&+31 24 08.2& 1955.310& 1996.279&  18.48& 17.88&  0.60&  17.70& 17.34&  0.36&     4.11&     3.35& 1.032& -26.4&  32.0&  0.96&            \nl
09 25 54.70&+40 04 14.4& 1953.178& 1996.279&  17.93& 17.12&  0.81&  17.74& 17.48&  0.26&     9.16&     9.30& 0.471& -25.3&  31.7&  1.08&            \nl
09 29 13.92&+37 57 43.1& 1954.022& 1996.279&  18.30& 17.72&  0.58&  17.59& 17.43&  0.16&    43.19&    43.43& 1.915& -28.1&  33.5&  1.84&            \nl
09 32 55.46&+28 40 36.6& 1952.164& 1996.304&  16.86& 16.66&  0.20&  16.69& 16.66&  0.03&     1.40&     0.94& 0.543& -26.1&  31.0&  0.06&            \nl
09 33 11.62&+35 16 27.4& 1954.022& 1996.304&  19.07& 17.84&  1.23&  18.23& 17.29&  0.94&     4.85&     4.97& 1.121& -26.7&  32.1&  1.01&            \nl
09 33 37.30&+28 45 32.2& 1952.164& 1996.304&  19.39& 18.78&  0.61&  17.86& 17.59&  0.27&   119.85&   120.71& 3.425& -28.4&  34.4&  2.61&            \nl
09 34 07.29&+35 42 37.6& 1954.022& 1996.304&  17.70& 17.71& -0.01&  17.11& 16.95&  0.16&     3.87&     5.75& 1.097& -26.7&  32.2&  1.03&core-jet 12s\nl
09 35 31.59&+35 41 00.9& 1954.022& 1996.304&  17.56& 17.34&  0.22&  16.95& 16.82&  0.13&     1.54&     2.10& 0.492& -25.2&  31.1&  0.52&            \nl
09 36 21.52&+39 21 31.8& 1953.178& 1997.184&  19.25& 17.93&  1.32&  17.80& 17.32&  0.48&    27.15&    28.51& 0.210& -22.7&  31.5&  1.93&            \nl
09 37 04.05&+29 37 04.3& 1952.164& 1996.304&  17.93& 17.72&  0.21&  17.62& 17.40&  0.22&     2.45&     2.65& 0.452& -24.6&  31.1&  0.77&FRII 160s   \nl
09 41 46.88&+35 30 35.9& 1954.022& 1996.304&  18.63& 18.16&  0.47&  17.66& 17.46&  0.20&     1.12&     0.93& 0.188& -22.2&  30.0&  0.62&            \nl
09 42 58.10&+40 10 14.2& 1953.178& 1997.184&  18.62& 17.51&  1.11&  17.65& 17.12&  0.53&     1.73&     2.02& 1.013& -26.8&  31.7&  0.50&            \nl
09 46 10.97&+32 23 25.8& 1952.164& 1997.186&  17.54& 17.25&  0.29&  17.13& 17.00&  0.13&     1.41&     1.56& 0.403& -24.8&  30.8&  0.36&            \nl
09 47 12.03&+34 51 19.9& 1955.279& 1996.304&  17.73& 17.76& -0.03&  16.94& 16.95& -0.01&     1.04&     1.20& 1.453& -27.4&  31.7&  0.33&            \nl
09 48 55.35&+40 39 44.8& 1953.178& 1996.304&  18.37& 17.49&  0.88&  18.00& 17.35&  0.65&  1439.97&  1537.07& 1.247& -27.3&  34.7&  3.35&            \nl
09 53 27.98&+32 25 51.7& 1955.279& 1996.304&  17.74& 17.05&  0.69&  17.05& 16.73&  0.32&   128.43&   131.85& 1.574& -28.3&  33.8&  2.08&            \nl
09 53 31.60&+39 17 30.3& 1953.178& 1996.304&  17.27& 17.12&  0.15&  16.79& 16.94& -0.15&     4.97&     4.45& 0.917& -26.9&  32.0&  0.75&            \nl
09 55 37.96&+33 35 04.0& 1955.279& 1996.304&  17.79& 17.37&  0.42&  17.44& 17.10&  0.34&    35.73&    36.61& 2.499& -29.1&  33.7&  1.58&            \nl
09 55 55.71&+35 16 52.5& 1955.279& 1996.304&  18.46& 17.56&  0.90&  17.59& 17.19&  0.40&     9.28&     9.71& 1.620& -27.8&  32.7&  1.15&            \nl
09 58 20.95&+32 24 02.6& 1955.279& 1996.304&  16.06& 15.81&  0.25&  15.88& 15.93& -0.05&  1204.19&  1228.71& 0.533& -26.9&  33.9&  2.66&            \nl
10 02 02.11&+37 42 24.3& 1955.279& 1996.307&  18.73& 18.10&  0.63&  19.30& 17.84&  1.46&     9.51&     8.19& 1.799& -27.6&  32.8&  1.34&            \nl
10 02 08.16&+34 53 53.7& 1955.279& 1996.307&  18.74& 18.36&  0.38&  17.05& 16.88&  0.17&     5.58&     5.30& 0.205& -22.2&  30.7&  1.39&            \nl
10 02 54.50&+32 40 39.6& 1955.279& 1996.307&  18.19& 17.11&  1.08&  17.47& 17.05&  0.42&     9.09&     9.00& 0.830& -26.7&  32.1&  1.01&            \nl
10 04 10.20&+37 28 48.7& 1955.279& 1996.515&  18.67& 17.97&  0.70&  18.06& 17.61&  0.45&     1.15&     0.96& 0.740& -25.5&  31.1&  0.47&            \nl
10 05 07.94&+34 14 24.5& 1955.279& 1996.515&  18.09& 17.11&  0.98&  16.80& 16.49&  0.31&     3.40&     2.66& 0.162& -22.9&  30.3&  0.69&            \nl
10 05 23.02&+40 58 34.1& 1953.288& 1996.515&  18.08& 17.64&  0.44&  16.95& 16.59&  0.36&     1.29&     1.07& 0.317& -23.9&  30.5&  0.45&            \nl
10 05 35.60&+32 51 26.2& 1955.279& 1996.518&  18.07& 17.28&  0.79&  17.26& 16.79&  0.47&     1.18&     0.66& 0.300& -24.1&  30.3&  0.24&            \nl
10 08 41.23&+36 23 19.1& 1955.279& 1996.518&  18.63& 18.03&  0.60&  17.45& 17.00&  0.45&    33.01&    34.77& 3.125& -29.0&  33.8&  1.79&            \nl
10 10 00.70&+30 03 21.6& 1955.373& 1996.304&* 17.39& 16.59&  0.80&  16.76& 16.20&  0.56&     1.54&     0.99& 0.255& -24.5&  30.4&  0.12&            \nl
10 10 27.58&+41 32 38.4& 1953.288& 1996.307&  16.45& 16.04&  0.41&  16.28& 16.03&  0.25&   258.74&   340.26& 0.613& -27.0&  33.5&  2.19&FRII 32s    \nl
10 15 05.77&+36 04 52.9& 1954.334& 1996.518&  18.50& 17.94&  0.56&  17.88& 17.55&  0.33&    29.51&    29.85& 0.846& -25.9&  32.7&  1.86&            \nl
10 18 58.57&+34 36 32.4& 1954.334& 1997.268&  17.41& 16.92&  0.49&  16.19& 16.04&  0.15&     2.27&     2.44& 0.109& -22.2&  29.8&  0.48&            \nl
10 20 41.12&+39 58 11.1& 1953.288& 1996.518&  19.45& 18.58&  0.87&  18.36& 17.63&  0.73&     2.71&     2.38& 0.826& -25.2&  31.6&  1.08&FRII 160s   \nl
10 21 17.55&+34 37 22.0& 1954.334& 1996.526&  17.54& 17.34&  0.20&  17.09& 17.19& -0.10&   317.10&   392.03& 1.407& -27.7&  34.2&  2.68&FRII 22s    \nl
10 21 56.57&+30 01 40.6& 1955.364& 1996.304&  18.29& 18.56& -0.27&  17.48& 17.21&  0.27&     1.43&     1.41& 3.115& -28.4&  32.4&  0.62&            \nl
10 22 30.33&+30 41 04.7& 1955.364& 1996.304&  17.58& 17.32&  0.26&  17.07& 17.01&  0.06&   907.01&   919.28& 1.320& -27.6&  34.5&  3.05&            \nl
10 22 37.38&+39 31 50.6& 1953.288& 1996.526&  17.25& 16.93&  0.32&  16.64& 16.32&  0.32&    41.28&    57.53& 0.607& -26.1&  32.7&  1.77&FRII 36s    \nl
10 23 11.52&+39 48 15.1& 1953.288& 1996.526&  18.17& 17.29&  0.88&  17.52& 16.81&  0.71&   807.88&  1078.67& 1.252& -27.5&  34.6&  3.12&            \nl
10 23 33.46&+39 53 12.5& 1953.288& 1997.186&  18.36& 17.67&  0.69&  17.80& 17.43&  0.37&    62.13&    65.13& 1.330& -27.3&  33.4&  2.04&            \nl
10 26 17.50&+30 36 42.7& 1955.364& 1997.186&  18.55& 18.12&  0.43&  17.47& 17.25&  0.22&     1.74&     1.82& 0.340& -23.6&  30.7&  0.79&            \nl
10 30 45.22&+25 55 22.4& 1955.359& 1997.268&* 17.47& 17.22&  0.25&  16.92& 16.21&  0.71&    47.56&    49.15& 0.691& -26.1&  32.7&  1.81&core-jet 12s\nl
10 30 59.11&+31 02 55.7& 1955.364& 1997.186&  16.29& 16.15&  0.14&  15.88& 15.68&  0.20&    58.74&    63.45& 0.178& -24.1&  31.7&  1.57&FRII 30s    \nl
10 33 03.77&+41 16 05.8& 1953.263& 1997.186&  19.04& 18.27&  0.77&  18.23& 17.49&  0.74&   406.68&   432.65& 1.120& -26.2&  34.1&  3.12&            \nl
10 33 59.48&+35 55 08.5& 1954.334& 1997.186&  17.12& 16.80&  0.32&  16.41& 16.35&  0.06&     2.06&     2.23& 0.169& -23.3&  30.2&  0.38&            \nl
10 36 48.48&+37 03 07.1& 1954.334& 1997.268&  18.47& 17.77&  0.70&  18.15& 18.24& -0.09&    26.23&    26.71& 1.592& -27.6&  33.2&  1.67&            \nl
10 38 48.12&+37 29 24.0& 1954.334& 1997.268&  17.50& 17.44&  0.06&  17.05& 17.51& -0.46&    17.47&    17.46& 0.730& -26.0&  32.3&  1.44&FRII 65s    \nl
10 38 59.61&+42 27 42.1& 1953.263& 1997.268&  18.49& 18.52& -0.03&  17.29& 17.32& -0.03&     2.76&     2.44& 0.220& -22.2&  30.5&  1.15&            \nl
10 52 50.06&+33 55 05.1& 1953.427& 1997.189&  17.67& 16.95&  0.72&  16.82& 16.77&  0.05&    12.81&    13.41& 1.408& -28.1&  32.8&  1.06&FRII 35s    \nl
11 02 11.87&+28 40 40.9& 1955.304& 1997.189&  17.46& 17.44&  0.02&  16.64& 16.81& -0.17&    46.17&    48.36& 1.741& -28.1&  33.5&  1.79&            \nl
11 03 13.34&+30 14 42.9& 1955.304& 1997.189&  18.18& 17.65&  0.53&  17.64& 17.90& -0.26&   107.68&   108.99& 0.387& -24.3&  32.6&  2.37&FRII 75s    \nl
11 06 05.69&+30 51 08.5& 1955.304& 1997.189&  17.45& 17.10&  0.35&  16.86& 16.80&  0.06&     1.69&     1.92& 1.638& -28.3&  32.0&  0.26&            \nl
11 10 40.21&+30 19 09.9& 1955.304& 1997.189&  18.97& 17.98&  0.99&  18.24& 17.86&  0.38&    20.08&    22.77& 1.522& -27.3&  33.1&  1.69&FRII 45s    \nl
11 22 41.49&+30 35 35.1& 1950.425& 1997.189&  17.65& 17.01&  0.64&  16.74& 16.74& -0.00&    10.06&    10.02& 1.809& -28.7&  32.8&  0.93&            \nl
11 31 09.29&+26 32 08.1& 1950.425& 1997.595&  16.96& 16.44&  0.52&  16.40& 15.46&  0.94&     2.00&     2.72& 0.243& -24.5&  30.6&  0.31&            \nl
12 02 43.54&+37 35 52.0& 1956.430& 1997.578&* 18.48& 17.54&  0.94&  17.77& 17.36&  0.41&    16.29&    17.24& 1.194& -27.1&  32.7&  1.43&            \nl
12 04 37.56&+28 51 25.1& 1955.373& 1997.578&  17.36& 17.07&  0.29&  16.64& 16.51&  0.13&     6.24&     5.57& 1.129& -27.5&  32.2&  0.80&            \nl
13 10 59.41&+32 33 34.6& 1950.449& 1996.279&  20.96& 16.79&  4.17&  20.56& 16.11&  4.45&   243.42&   245.97& 1.650& -28.7&  34.2&  2.24&            \nl
13 11 31.98&+37 34 08.1& 1950.449& 1996.304&  17.31& 17.11&  0.20&  17.11& 16.83&  0.28&     1.20&     1.53& 1.272& -27.7&  31.7&  0.19&            \nl
13 12 17.75&+35 15 20.6& 1950.449& 1996.304&  15.55& 15.66& -0.11&  15.05& 14.86&  0.19&    43.92&    44.72& 0.183& -24.6&  31.5&  1.22&            \nl
13 14 23.14&+29 10 01.7& 1955.367& 1996.304&* 18.73& 18.30&  0.43&  17.90& 17.52&  0.38&     7.53&     7.22& 1.094& -26.2&  32.3&  1.38&            \nl
13 26 29.06&+36 23 34.3& 1950.452& 1996.696&  18.10& 17.60&  0.50&  17.63& 17.23&  0.40&     1.75&     1.07& 0.389& -24.4&  30.8&  0.56&            \nl
13 31 08.31&+30 30 32.9& 1950.507& 1996.304&  17.49& 17.38&  0.11&  17.14& 17.11&  0.03& 14777.96& 15023.95& 0.848& -26.5&  35.4&  4.34&            \nl
13 33 09.74&+36 20 30.4& 1950.452& 1996.701&  17.95& 17.46&  0.49&  18.53& 17.38&  1.15&    16.63&    17.50& 0.679& -25.9&  32.3&  1.46&core-jet 7s \nl
13 34 03.81&+37 01 04.0& 1950.452& 1996.696&  18.38& 17.65&  0.73&  17.45& 17.05&  0.40&    13.17&    13.31& 1.765& -28.0&  32.9&  1.31&            \nl
13 41 59.89&+37 07 10.5& 1950.452& 1996.701&  17.16& 16.67&  0.49&  17.84& 16.61&  1.23&    24.50&    25.42& 1.106& -27.8&  32.8&  1.26&            \nl
13 42 54.40&+28 28 06.1& 1950.507& 1996.515&  18.50& 18.20&  0.30&  17.83& 17.77&  0.06&    65.57&    68.41& 1.040& -26.1&  33.2&  2.30&FRII 35s    \nl
13 43 00.18&+28 44 07.3& 1950.507& 1996.515&  17.23& 17.00&  0.23&  16.77& 16.43&  0.34&   217.37&   246.65& 0.905& -27.0&  33.7&  2.40&            \nl
13 48 04.37&+28 40 25.6& 1958.373& 1996.515&  18.00& 17.89&  0.11&  17.77& 17.60&  0.17&    72.45&    78.07& 2.464& -28.5&  34.0&  2.12&            \nl
13 48 20.89&+30 20 05.3& 1958.373& 1996.515&  17.36& 18.09& -0.73&  16.66& 17.57& -0.91&    22.49&    23.53& 1.865& -27.7&  33.2&  1.72&core-jet 15s\nl
13 53 26.08&+36 20 49.6& 1950.452& 1996.518&  16.30& 15.90&  0.40&  15.97& 15.63&  0.34&     1.70&     1.57& 0.286& -25.4&  30.5& -0.12&            \nl
13 55 29.94&+29 30 58.9& 1958.373& 1996.518&  18.25& 18.47& -0.22&  17.43& 17.33&  0.10&     1.74&     1.54& 0.302& -23.0&  30.6&  0.91&            \nl
14 10 36.79&+29 55 50.7& 1958.373& 1996.518&  18.61& 17.69&  0.92&  18.14& 17.76&  0.38&     4.00&     4.09& 0.570& -25.2&  31.5&  0.93&            \nl
14 15 28.45&+37 06 21.7& 1950.447& 1996.518&* 18.20& 18.37& -0.17&  17.76& 17.45&  0.31&   406.30&   409.69& 2.360& -27.9&  34.7&  3.04&            \nl
14 23 26.09&+32 52 20.6& 1950.466& 1997.496&  17.25& 16.67&  0.58&  16.16& 16.23& -0.07&     8.71&     8.16& 1.903& -29.1&  32.8&  0.72&            \nl
14 25 50.71&+24 04 03.5& 1950.299& 1997.496&* 17.93& 17.78&  0.15&  17.13& 17.14& -0.01&   121.14&   320.88& 0.654& -25.4&  33.5&  2.85&FRII 22s    \nl
14 31 20.55&+39 52 41.8& 1950.532& 1996.523&* 16.47& 16.68& -0.21&  15.87& 15.69&  0.18&   207.87&   210.04& 1.217& -28.0&  33.8&  2.17&            \nl
14 37 19.19&+38 04 38.1& 1950.466& 1996.690&  18.24& 17.57&  0.67&  17.75& 17.41&  0.34&    34.25&    35.08& 1.043& -26.8&  32.9&  1.76&            \nl
14 37 56.48&+35 19 36.8& 1950.466& 1996.526&  18.16& 17.85&  0.31&  17.95& 17.42&  0.53&    16.53&    53.48& 0.537& -24.9&  32.5&  2.12&FRII 20s    \nl
14 55 43.46&+30 03 22.6& 1955.364& 1996.282&* 17.39& 16.69&  0.70&  16.94& 16.21&  0.73&     1.80&     1.41& 0.626& -26.4&  31.2&  0.17&            \nl
15 14 43.03&+36 50 50.4& 1956.499& 1996.282&  16.49& 15.78&  0.71&  16.30& 15.87&  0.43&    48.97&    70.99& 0.369& -26.1&  32.3&  1.44&FRII 60s    \nl
15 17 28.49&+28 56 15.8& 1954.564& 1996.282&  18.65& 17.86&  0.79&  17.53& 17.40&  0.13&     1.39&     0.84& 0.208& -22.7&  30.1&  0.59&            \nl
15 19 36.16&+28 38 27.0& 1954.564& 1996.282&  17.22& 17.37& -0.15&  16.61& 16.75& -0.14&     2.08&     1.99& 0.269& -23.8&  30.5&  0.56&            \nl
15 23 14.50&+37 59 28.4& 1950.447& 1996.690&  18.29& 18.44& -0.15&  17.43& 17.64& -0.21&     1.67&     1.83& 1.344& -26.5&  31.9&  0.80&            \nl
15 25 23.58&+42 01 17.1& 1955.329& 1996.523&  18.43& 18.47& -0.04&  17.74& 17.64&  0.10&   104.19&   122.19& 1.198& -26.2&  33.6&  2.65&core-jet 9s \nl
15 29 18.04&+32 48 42.3& 1950.447& 1996.690&  17.70& 17.52&  0.18&  17.05& 16.99&  0.06&    28.91&    29.62& 1.652& -27.9&  33.2&  1.61&            \nl
15 30 45.79&+38 39 52.7& 1955.329& 1996.282&  18.54& 17.91&  0.63&  17.82& 17.36&  0.46&     2.28&     2.28& 2.035& -28.0&  32.3&  0.62&            \nl
15 37 30.98&+33 58 36.1& 1950.447& 1996.696&  18.34& 17.74&  0.60&  17.22& 17.43& -0.21&     5.27&     5.58& 1.025& -26.6&  32.1&  1.03&            \nl
15 40 42.98&+41 38 16.0& 1955.329& 1996.696&  17.70& 17.93& -0.23&  17.28& 17.15&  0.13&    17.67&    18.28& 2.506& -28.5&  33.4&  1.51&            \nl
15 42 27.03&+29 42 02.4& 1950.460& 1996.282&  18.55& 18.10&  0.45&  17.88& 17.64&  0.24&     1.54&     1.35& 0.457& -24.3&  30.9&  0.69&            \nl
15 42 41.12&+29 34 28.9& 1950.460& 1996.282&  18.68& 18.19&  0.49&  17.99& 17.90&  0.09&     1.58&     1.81& 1.098& -26.3&  31.7&  0.72&            \nl
15 43 48.62&+40 13 24.8& 1955.329& 1997.578&  17.91& 17.72&  0.19&  17.14& 16.79&  0.35&     3.24&     2.75& 0.318& -23.8&  30.9&  0.88&            \nl
15 44 05.61&+32 40 49.1& 1950.447& 1997.578&  17.70& 17.92& -0.22&  17.04& 17.52& -0.48&   210.79&   213.88& 1.047& -26.4&  33.7&  2.69&            \nl
15 45 12.93&+30 05 08.0& 1950.460& 1996.523&  18.50& 19.24& -0.74&  17.71& 17.83& -0.12&     6.73&     6.26& 0.322& -22.3&  31.2&  1.81&            \nl
15 48 17.94&+35 11 27.5& 1950.359& 1996.523&  18.26& 18.04&  0.22&  17.92& 17.70&  0.22&   140.94&   141.51& 0.479& -24.5&  32.9&  2.63&            \nl
15 54 29.39&+30 01 19.0& 1950.460& 1996.523&* 18.38& 18.08&  0.30&  17.75& 17.18&  0.57&    40.04&    41.22& 2.686& -28.6&  33.8&  1.91&            \nl
15 58 55.20&+33 23 19.4& 1950.359& 1996.523&  18.02& 17.96&  0.06&  16.94& 16.94& -0.00&   142.96&   144.89& 1.655& -27.5&  33.9&  2.48&            \nl
16 02 57.37&+30 38 51.6& 1950.512& 1996.523&  18.05& 18.24& -0.19&  17.60& 17.81& -0.21&     1.23&     1.39& 0.809& -25.5&  31.3&  0.65&            \nl
16 03 54.15&+30 02 08.5& 1950.512& 1996.523&  18.17& 17.88&  0.29&  17.72& 17.50&  0.22&    53.69&    54.18& 2.026& -28.1&  33.7&  1.99&            \nl
16 06 48.11&+29 10 48.0& 1950.512& 1996.523&  18.82& 18.42&  0.40&  17.67& 17.40&  0.27&     1.26&     0.74& 0.328& -23.2&  30.5&  0.75&            \nl
16 11 23.26&+29 59 47.2& 1950.512& 1996.526&  18.50& 18.56& -0.06&  17.85& 17.75&  0.10&     1.69&     1.90& 1.711& -27.0&  32.1&  0.83&            \nl
16 13 41.09&+34 12 48.3& 1954.564& 1996.526&  18.11& 17.74&  0.37&  17.23& 17.53& -0.30&  3532.04&  3605.54& 1.404& -27.3&  35.2&  3.81&core-jet 11s\nl
16 14 36.84&+28 39 06.2& 1950.512& 1996.523&  19.39& 19.05&  0.34&  17.80& 17.51&  0.29&     2.22&     1.31& 0.316& -22.5&  30.7&  1.25&            \nl
16 19 02.43&+30 30 51.4& 1950.512& 1996.526&  17.24& 17.29& -0.05&  16.56& 16.46&  0.10&    66.82&    67.58& 1.284& -27.5&  33.4&  1.91&            \nl
16 23 07.39&+30 04 06.2& 1954.567& 1996.526&  18.08& 17.29&  0.79&  17.32& 16.95&  0.37&     1.65&     1.04& 1.165& -27.3&  31.7&  0.31&            \nl
16 23 19.92&+41 17 02.8& 1953.460& 1997.496&  17.46& 17.32&  0.14&  16.64& 16.67& -0.03&     3.23&     2.98& 1.618& -28.1&  32.3&  0.57&            \nl
16 23 30.56&+35 59 32.8& 1954.564& 1996.526&  18.55& 17.91&  0.64&  17.90& 17.59&  0.31&   259.58&   266.83& 0.867& -26.0&  33.6&  2.80&            \nl
16 24 22.02&+39 24 41.5& 1953.460& 1996.526&  18.94& 18.31&  0.63&  18.26& 17.76&  0.50&   132.32&   144.52& 1.117& -26.2&  33.6&  2.66&FRII 22s    \nl
16 25 48.78&+26 46 58.5& 1954.567& 1997.595&  17.40& 17.09&  0.31&  17.01& 17.05& -0.04&    10.12&     9.69& 2.521& -29.4&  33.1&  0.91&            \nl
16 26 59.26&+30 15 34.9& 1954.567& 1996.526&  18.36& 18.17&  0.19&  17.55& 17.19&  0.36&     5.34&     5.42& 1.582& -27.2&  32.5&  1.14&            \nl
16 29 01.32&+40 08 00.0& 1953.460& 1996.526&  17.67& 18.11& -0.44&  17.25& 17.86& -0.61&    11.97&    11.94& 0.271& -23.1&  31.3&  1.61&            \nl
16 30 20.76&+37 56 56.2& 1954.564& 1996.690&  16.59& 17.31& -0.72&  16.43& 17.45& -1.02&    20.00&    21.73& 0.395& -24.7&  31.9&  1.53&            \nl
16 33 02.11&+39 24 27.5& 1953.460& 1996.690&  17.17& 16.86&  0.31&  16.56& 15.97&  0.59&    41.29&    54.38& 1.021& -27.4&  33.1&  1.67&FRII 20s    \nl
16 33 48.99&+31 34 11.8& 1954.567& 1996.690&  17.38& 17.72& -0.34&  16.87& 17.27& -0.40&     1.79&     2.07& 1.516& -27.5&  32.0&  0.55&            \nl
16 34 02.97&+39 00 00.7& 1953.460& 1996.690&  18.91& 18.38&  0.53&  18.22& 18.03&  0.19&   915.40&   931.91& 1.085& -26.1&  34.4&  3.51&            \nl
16 34 12.78&+32 03 35.2& 1954.567& 1996.515&  17.61& 17.26&  0.35&  17.23& 16.77&  0.46&   175.25&   176.52& 2.341& -29.0&  34.3&  2.23&            \nl
16 35 15.46&+38 08 04.1& 1954.564& 1996.690&  17.08& 17.83& -0.75&  16.33& 17.61& -1.28&  2653.87&  2694.06& 1.811& -27.8&  35.3&  3.68&            \nl
16 37 09.33&+41 40 30.4& 1953.460& 1996.690&  16.95& 17.38& -0.43&  16.64& 17.05& -0.41&     7.37&     7.11& 0.765& -26.2&  32.0&  1.04&            \nl
16 39 31.75&+39 08 45.7& 1953.460& 1996.690&  18.36& 17.91&  0.45&  17.46& 17.28&  0.18&     1.27&     0.72& 0.143& -21.8&  29.8&  0.58&            \nl
16 40 29.62&+39 46 46.2& 1953.460& 1996.690&  19.37& 18.64&  0.73&  18.59& 17.96&  0.63&  1088.22&  1108.55& 1.670& -26.8&  34.8&  3.63&            \nl
16 41 54.24&+40 00 33.0& 1953.460& 1997.595&  18.10& 17.99&  0.11&  17.63& 17.68& -0.05&     6.89&     5.06& 1.003& -26.3&  32.2&  1.23&            \nl
16 42 18.96&+31 54 33.9& 1954.567& 1996.611&  18.55& 18.21&  0.34&  17.93& 17.78&  0.15&    23.08&    24.01& 1.263& -26.6&  32.9&  1.83&            \nl
16 42 58.78&+39 48 37.1& 1953.460& 1996.690&  17.62& 16.19&  1.43&  17.12& 15.69&  1.43&  6050.06&  6598.61& 0.595& -26.8&  34.7&  3.54&core-jet 12s\nl
16 43 47.87&+30 11 08.6& 1954.567& 1996.699&  17.67& 17.45&  0.22&  17.00& 16.77&  0.23&     1.42&     1.38& 1.380& -27.6&  31.8&  0.29&            \nl
16 50 05.45&+41 40 32.3& 1953.460& 1996.690&  17.85& 17.18&  0.67&  17.46& 17.33&  0.13&   170.28&   185.19& 0.584& -25.8&  33.2&  2.38&complex     \nl
16 52 52.62&+36 00 57.3& 1954.482& 1996.696&  18.53& 18.80& -0.27&  17.68& 17.71& -0.03&     1.36&     2.22& 0.281& -22.5&  30.6&  1.16&            \nl
16 52 55.92&+31 23 43.5& 1950.466& 1996.699&  17.79& 17.94& -0.15&  17.51& 17.87& -0.36&     1.16&     1.47& 0.590& -25.0&  31.1&  0.58&            \nl
16 55 00.24&+30 30 40.0& 1950.466& 1996.699&  18.21& 17.63&  0.58&  17.66& 17.34&  0.32&     3.20&     2.86& 0.405& -24.5&  31.1&  0.83&            \nl
16 55 57.40&+32 18 05.3& 1950.466& 1996.699&  18.34& 18.19&  0.15&  17.51& 17.42&  0.09&     1.10&     1.19& 1.327& -26.7&  31.7&  0.51&            \nl
16 57 22.12&+39 55 51.7& 1954.597& 1996.699&  17.89& 17.83&  0.06&  17.50& 17.45&  0.05&     1.19&     0.77& 0.579& -25.1&  31.0&  0.45&            \nl
16 59 24.14&+26 29 37.0& 1955.329& 1997.595&* 18.55& 18.21&  0.34&  18.07& 17.90&  0.17&   391.07&   538.46& 0.794& -25.5&  33.9&  3.23&core-jet 14s\nl
16 59 31.91&+37 35 28.4& 1954.482& 1996.699&  17.45& 17.34&  0.11&  17.14& 17.41& -0.27&    18.31&    18.57& 0.775& -26.3&  32.4&  1.42&            \nl
17 02 31.01&+32 47 20.6& 1954.482& 1996.526&  16.21& 16.09&  0.12&  15.71& 16.03& -0.32&     1.82&     1.52& 0.163& -23.9&  30.0&  0.01&            \nl
17 06 48.07&+32 14 22.7& 1950.466& 1996.690&  16.87& 17.06& -0.19&  16.27& 16.46& -0.19&    36.32&    38.71& 1.070& -27.3&  33.0&  1.60&FRII 56s    \nl
17 08 23.08&+41 23 09.4& 1954.597& 1996.699&  17.36& 17.26&  0.10&  17.08& 16.85&  0.23&     1.23&     0.94& 0.837& -26.5&  31.3&  0.21&            \nl
17 10 13.50&+33 44 03.3& 1954.592& 1997.496&  16.45& 15.93&  0.52&  15.64& 15.25&  0.39&     4.05&     4.60& 0.208& -24.7&  30.7&  0.34&            \nl
17 13 04.48&+35 23 33.8& 1954.592& 1996.526&  16.95& 16.80&  0.15&  15.99& 15.85&  0.14&    11.13&    11.24& 0.083& -21.7&  30.2&  1.10&            \nl
17 16 01.95&+31 12 13.7& 1950.540& 1996.526&  15.93& 15.39&  0.54&  15.30& 14.61&  0.69&     2.68&     2.42& 0.110& -23.8&  29.9& -0.09&            \nl
17 16 54.19&+30 27 01.4& 1950.540& 1997.496&  17.45& 17.07&  0.38&  16.87& 16.74&  0.13&     4.25&     3.96& 0.751& -26.5&  31.7&  0.68&            \nl
17 19 34.17&+25 10 58.7& 1951.589& 1997.595&  16.36& 16.57& -0.21&  15.74& 15.76& -0.02&    12.67&    13.83& 0.579& -26.4&  32.0&  1.01&            \nl
17 20 07.69&+36 54 39.2& 1954.592& 1997.595&  17.75& 17.82& -0.07&  16.91& 16.87&  0.04&     3.42&     3.34& 0.332& -23.8&  30.9&  0.94&            \nl
17 21 09.50&+35 42 16.2& 1954.592& 1996.688&  19.12& 17.14&  1.98&  17.93& 16.58&  1.35&   386.51&   392.99& 0.283& -24.1&  32.9&  2.74&FRII 130s   \nl
17 23 20.79&+34 17 57.6& 1954.592& 1996.688&  15.77& 15.16&  0.61&  15.48& 14.96&  0.52&   438.57&   441.20& 0.205& -25.4&  32.6&  2.01&FRII 200s   \nl
17 23 54.30&+37 48 41.1& 1954.592& 1996.690&  17.72& 18.43& -0.71&  17.13& 17.64& -0.51&     1.70&     1.72& 0.828& -25.4&  31.4&  0.82&            \nl
17 26 32.71&+39 57 02.1& 1953.529& 1996.688&  18.40& 18.49& -0.09&  18.06& 18.00&  0.06&   475.49&   496.89& 0.656& -24.7&  33.7&  3.33&            \nl
17 26 35.10&+32 13 23.0& 1950.540& 1996.699&  17.99& 17.78&  0.21&  17.36& 17.24&  0.12&   120.94&   123.67& 1.094& -26.7&  33.5&  2.39&core-jet 16s\nl
17 28 59.09&+38 38 26.0& 1954.592& 1996.688&  17.06& 16.86&  0.20&  16.53& 16.76& -0.23&   240.27&   245.41& 1.391& -28.2&  34.0&  2.29&            \nl
17 34 03.52&+40 37 54.0& 1953.529& 1997.595&  17.27& 16.50&  0.77&  16.38& 15.91&  0.47&     1.07&     2.53& 0.356& -25.3&  30.9&  0.28&            \nl
21 35 13.10&-00 52 43.8& 1954.589& 1996.688&  18.49& 17.96&  0.53&  17.88& 17.66&  0.22&     1.84&     1.32& 1.660& -27.5&  32.0&  0.58&FRII 55s    \nl
21 36 38.60&+00 41 54.5& 1954.589& 1996.688&  17.20& 17.25& -0.05&  16.72& 16.65&  0.07&  3546.71&  3712.01& 1.930& -28.6&  35.5&  3.58&            \nl
21 37 48.50&+00 12 20.5& 1954.589& 1996.688&  18.40& 18.14&  0.26&  17.92& 17.72&  0.20&    36.02&    39.56& 1.666& -27.3&  33.4&  1.98&            \nl
21 59 24.09&+01 13 05.3& 1954.652& 1996.701&  17.05& 16.57&  0.48&  16.24& 15.69&  0.55&     1.45&     2.10& 1.000& -27.7&  31.7&  0.14&            \nl
22 01 03.12&-00 52 59.6& 1954.652& 1996.701&  17.34& 16.73&  0.61&  16.24& 16.10&  0.14&     1.47&     1.58& 0.213& -23.9&  30.2&  0.19&            \nl
22 03 14.81&-01 31 23.4& 1954.652& 1996.701&  16.78& 16.51&  0.27&  16.46& 16.24&  0.22&    14.21&    14.25& 0.650& -26.7&  32.1&  0.99&            \nl
22 03 55.60&+00 55 16.7& 1954.652& 1996.718&  18.65& 17.70&  0.95&  18.08& 17.31&  0.77&     2.75&     2.49& 0.720& -25.8&  31.5&  0.75&            \nl
22 06 25.99&-01 52 01.1& 1954.652& 1996.718&  18.35& 18.14&  0.21&  17.18& 17.17&  0.01&     2.60&     2.97& 1.110& -26.3&  31.9&  0.91&            \nl
22 18 06.66&+00 52 23.9& 1954.652& 1996.929&  16.56& 17.28& -0.72&  16.12& 16.86& -0.74&    53.42&    54.77& 1.270& -27.5&  33.3&  1.82&            \nl
23 44 03.12&+00 38 03.9& 1951.989& 1996.696&  17.27& 16.68&  0.59&  16.82& 16.58&  0.24&     1.57&     1.65& 1.230& -28.1&  31.7&  0.06&            \nl
23 55 20.58&+00 07 48.0& 1951.989& 1996.934&  18.56& 18.22&  0.34&  17.78& 17.83& -0.05&    17.72&    18.92& 1.070& -26.2&  32.7&  1.75&core-jet 8s \nl
\enddata
%
\tablecomments{Descriptions of table columns:\\
Cols 1--2: FIRST radio positions (J2000).\\
Col 3: Date of POSS-I observation.\\
Col 4: Date of CCD observation.\\
Cols 5--10: Extinction-corrected CCD magnitudes (B, R), POSS-I magnitudes
(O, E) and colors.  An asterisk preceding B indicates sources with too few
CCD calibration stars to recalibrate the POSS-I magnitudes (see text.)\\
Cols 11-12: 20~cm peak (col.~11) and integrated (col.~12) flux densities
in mJy from the FIRST catalog.\\
Col 13: Redshift.\\
Col 14: Absolute B magnitude.\\
Col 15: Log of radio luminosity (erg cm$^{-2}$ s$^{-1}$ Hz$^{-1}$)
at a rest-frame frequency of 5~GHz, assuming spectral index
$\alpha = -0.5$ ($F_\nu \propto \nu^\alpha$).\\
Col 16: $K$-corrected 5~GHz radio to 2500~\AA\ optical luminosity ratio
using the definition of Stocke et al.\ (1992).\\
Col 17: Radio morphology.  Categories are FRII (double, with approximate
separation), core-jet (with approximate jet length), and complex.  Objects
with no notation are symmetrical, isolated sources.
}
\end{deluxetable}
\endgroup

\end{document}